\documentclass[12pt,preprint]{aastex}
\usepackage{graphicx}
\usepackage{subfigure}
\usepackage{epsfig}
\usepackage{rotating}
\usepackage{floatrow}
\DeclareFloatFont{scriptsize}{\scriptsize}% "scriptsize" is defined by floatrow, "tiny" not
\newcommand{\g}{\Gamma}
\newcommand{\s}{\sigma}

\begin{document}

\title{Photosphere emission from a hybrid relativistic outflow with arbitrary dimensionless entropy and 
magnetization in GRBs}

\author{He Gao, Bing Zhang}
\affil{Department of Physics and Astronomy, University of Nevada Las Vegas, NV 89154, USA \\
gaohe@physics.unlv.edu;~zhang@physics.unlv.edu}

\begin{abstract}

In view of the recent \emph{Fermi} observations of GRB prompt emission spectra, we develop a theory of photosphere emission of a hybrid relativistic outflow with a hot fireball component (defined by dimensionless entropy $\eta$) and a cold Poynting-flux component (defined by magnetization $\sigma_0$ at the central engine).  We consider the scenarios both without and with sub-photospheric magnetic dissipations. Based on a simplified toy model of jet dynamics, we develop two approaches: a ``bottom-up'' approach to predict the temperature (for a non-dissipative photosphere) and luminosity of the photosphere emission and its relative brightness for a given pair of $(\eta,\sigma_0)$; and a ``top-down'' approach to diagnose central engine parameters ($\eta$ and $\sigma_0$) based on the observed quasi-thermal photosphere emission properties. We show that a variety of observed GRB prompt emission spectra with different degrees of photosphere thermal emission can be reproduced by varying $\eta$ and $\sigma_0$ within the non-dissipative photosphere scenario. In order to reproduce the observed spectra, the outflows of most GRBs need to have a significant $\sigma$, both at the central engine, and at the photosphere. The $\sigma$ value at $10^{15}$ cm from the central engine (a possible non-thermal emission site) is usually also greater than unity, so that internal-collision-induced magnetic reconnection and turbulence (ICMART) may be the mechanism to power the non-thermal emission. We apply our top-down approach to GRB 110721A, and find that the temporal evolution behavior of its blackbody component can be well interpreted with a time-varying $(\eta,\sigma_0)$ at the central engine, instead of invoking a varying engine base size $r_0$ as proposed by previous authors. 
\end{abstract}

\section{Introduction}

After decades of investigations, the origin of gamma-ray burst (GRB) prompt emission is still poorly understood. The main obstacle in front of theorists is one fundamental question: \emph{What is the composition of GRB jets? } 

In the early picture discussed by \cite{paczynski86} and \cite{goodman86}, GRB outflows originate from an initially hot ``fireball" composed of photons and electron/positron pairs.
The emergent spectrum from the fireball photosphere is a modified blackbody, which is inconsistent with the typical observed spectrum, characterized by a smoothly-joint broken power-law function known as the ``Band" function \citep{band93}. Observationally, the typical low- and high-energy photon indices of the Band function are distributed around $\alpha~\sim -1$ and $\beta~\sim -2.2$, respectively \citep{preece00}, which disfavor a simplest fireball photosphere model.  \cite{shemi90} showed that when a small amount of baryons are added to the fireball, a significant fraction of the initial fireball thermal energy is converted to the kinetic energy of the outflow, after an initially rapid acceleration phase under fireball's thermal pressure \citep{meszaros93,piran93}. In order to produce non-thermal photons, the kinetic energy of the outflow needs to be dissipated, either in the external shock \citep{rees92,meszarosrees93} or internal shocks \citep[hereafter IS,][]{rees94}. Synchrotron (and possibly also synchrotron self-Compton) radiation by the relativistic electrons accelerated in these shocks give rise to the observed non-thermal $\gamma$-ray emission \citep{meszaros94,tavani96,daigne98,lloyd00}. Within such a ``fireball shock'' model, the observed spectrum is expected to be the superposition of two components: a non-thermal component from the IS in the optically thin region, and a quasi-thermal component from the fireball photosphere \citep{meszaros00,meszaros02,pe'er06}.  

Before {\em Fermi}, it has been claimed that the spectra of some BATSE GRBs can be fit with the superposition of a blackbody component and an underlying power law 
\cite[e.g.][]{ghirlanda02,ghirlanda03,ryde05,ryde09}. Due to the narrowness of the energy band, one was not able to exclude other models, so as to firmly establish the thermal model. Most BATSE GRBs, on the other hand, have a dominant Band function component for both time-integrated and time-resolved spectra. There are several competing models to interpret the peak energy $E_p$ in the pre-{\em Fermi} era \citep{zhang02}. If it is the synchrotron peak from the IS, then the absence of the photospheric emission component would imply that GRB jets are magnetically dominated \citep{daigne02,zhang02}. Within this scenario, the GRB radiation is powered by dissipation of the magnetic field energy in the ejecta \citep{usov94,thompson94,meszaros97,lyutikov03}. Alternatively, the Band function itself may be emission from a dissipative photosphere \citep{rees05,giannios08,beloborodov10,ioka10,lazzati10,pe'er11,vurm11,giannios12,lundman13}.

Having both Gamma-ray Burst Monitor (GBM, \citealt{meegan09}) and Large Area Telescope (LAT, \citealt{atwood09}) on board, {\em Fermi} opened the spectral window to cover 6-7 orders of magnitude in energy, allowing a close investigation of various spectral components in the GRB prompt emission spectra. The first bright LAT GRB, GRB 080916C, showed nearly featureless Band spectra in 5 different time bins over 6-7 orders of magnitude in energy \citep{abdo09a,zhang11}. Assuming the standard fireball shock model and using parameters derived from the observational data, \cite{zhang09} showed that a quasi-thermal photosphere component would greatly outshine the non-thermal component if GRB 080916C is a fireball. The non-detection of such a bright photosphere component allowed \cite{zhang09} to claim that the jet of GRB 080916C is Poynting-flux-dominated, and a lower limit of $\sigma$ was set at the photosphere\footnote{A recent more detailed investigation (S. Guiriec et al. 2014, in preparation) revealed a weak photosphere component in GRB 080916C. Its peak flux is roughly at the flux level that \cite{zhang09} used to derive the lower limit of $\sigma$. As a result, the conclusion of \cite{zhang09} that GRB 080916C is a Poynting flux dominated remains valid, with the derived lower limit of $\sigma$ replaced by the real value of $\sigma$.}. In order to interpret the non-thermal emission in a Poynting flux dominated outflow, \cite{zhangyan11} proposed an ``internal-collision-induced magnetic reconnection and turbulence'' (ICMART) model for GRBs.  

The quasi-thermal component predicted by the fireball model was later observed in some {\em Fermi} GRBs. GRB 090902B \citep{abdo09b} is the most prominant one, which shows a narrow Band function component superposed on an underlying power law component in the time integrated spectrum. When a time-resolved spectral analysis was carried out, a multi-color blackbody \citep{ryde10} or even a blackbody \citep{zhang11} component was revealed. This component is well interpreted as photosphere emission from a fireball \citep{pe'er12}. Later, several more GRBs have a thermal spectral component identified in the time-integrated and time-resolved spectra, but the component is sub-dominant (e.g. GRB 100724B, \citealt{guiriec11}; GRB 110721A, \citealt{axelsson12}; and the short GRB 120323A \citealt{guiriec13}). Since these thermal components are superposed on a Band component, it suggests that the Band emission component is not the modified thermal emission from the photosphere\footnote{The photosphere model suffers from the difficulty to account for the low-energy photon index of the Band function, and may not account for the observed hard-to-soft $E_p$ evolution pattern as observed in many GRBs \citep{deng14}.}. Rather, the Band component likely originates from an optically thin region, e.g. the IS or the ICMART site. The fact that GRB 110721A has a 15-MeV $E_p$ at early epochs disfavors a non-dissipative photospheric origin of the Band component, since it exceeds the maximum temperature a non-dissipative photosphere can reach for the observed luminosity (\citealt{zhang12}, see also \citealt{veres12b}).

The available data suggest that the photosphere emission in GRBs has diverse properties. While in rare cases (e.g. GRB 090902B) it can be the dominant emission component, in many more cases, it is either a sub-dominant component, or non-detectable. A plausible interpretation \citep[e.g.][]{zhang11b} would be that GRBs have a diverse jet composition. While some are more matter-dominated (which resemble traditional fireballs), many others have more magnetized ejecta with a range of magnetization degree ($\sigma$) at the central engine, the photosphere, and the non-thermal emission site. Two key parameters are the dimensionless entropy $\eta$ and magnetization $\sigma_0$ at the central engine. If $\eta$ is large and $\sigma_0 \ll 1$, one gets a hot fireball with a dominant photosphere component (e.g. in GRB 090902B). If $\eta$ is smaller while $\sigma_0$ is larger, the thermal emission is suppressed, so that the photosphere emission component is sub-dominant (e.g. in GRB 110721A). Finally, if $\eta$ is close to unity while $\sigma_0$ is extremely large, the photosphere component is completely suppressed. This is an attractive possibility. However, so far no theory of photosphere emission of such a hybrid outflow has been developed in detail.

Within the framework of the fireball shock model, \cite{pe'er07} proposed a method to infer central engine parameters using observed data. With the measured temperature and flux of an identified thermal component in the spectrum, along with a flux ratio between thermal and non-thermal components, one may infer the size of the jet at the base of the outflow, $r_0$, and the dimensionless entropy of the outflow, $\eta=L_w/\dot{M}c^2$ (which is also the bulk Lorentz factor of the outflow, if the photosphere radius is greater than the fireball coasting radius). Some authors have applied this method to some \emph{Fermi} GRBs \citep{iyyani13,preece14,ghirlanda13}. The derived central engine parameters are sometimes ad hoc or inconsistent. For instance, the analyses for GRB 110721A \citep{iyyani13} and for GRB 130427A \citep{preece14} led to a curious conclusion that the bulk Lorentz factor of the outflow of different layers are decreasing with time. This would lead to no, or at most very inefficient, internal shock emission. Yet both bursts have dominant non-thermal emission. More curiously, the data of GRB 110721A \citep{iyyani13} require that $r_0$ is rapidly varying with time by 2-3 orders of magnitudes. This is hard to imagine given the well believed paradigm of GRB central engine: If the engine is naked, the size of the engine (a hyper-accreting black hole or a millisecond magnetar) is around $r_0 \sim 10^7$ cm; if an extended envelope of a collapsar progenitor is considered, the fireball may be ``re-born'', with $r_0 \sim R_* \theta_j \sim 10^{9} R_{*,10} \theta_{j,-1}$ cm (where $R_*$ is the size of the progenitor star, and $\theta_j$ is the jet opening angle). If one considers the depletion of the envelope, $r_0$ should decrease with time. However, \cite{iyyani13} showed that $r_0$ increases from $10^6$ cm to $10^8$ cm early on, and then decreases mildly after 2 seconds. These absurd conclusions suggest that the starting point of the analysis, i.e. the assumption of a pure fireball model, might not be valid. It is interesting to see whether a hybrid ejecta photosphere model may solve the problem. Incidentally, \cite{ghirlanda13} analyzed another burst GRB 100507 using the fireball framework \citep{pe'er07}, but found that the derived $r_0$ remains constant and reasonable. The jet composition of that burst may be more close to a fireball. It would be interesting to see whether a general theoretical framework can be established, which may be reduced to the standard fireball framework when $\sigma_0 < 1$.

In this paper, we develop a theory of photosphere emission from a hybrid relativistic outflow with an arbitrary dimensionless entropy $\eta$ and magnetization $\sigma_0$ at the central engine based on a simplified toy model of the dynamical evolution of a hybrid jet. In section 2, we describe the set-up of the problem, and introduce an approximate analytical description of the dynamical evolution of the hybrid system. In section 3,  we develop a ``bottom-up'' approach by deriving the photosphere properties for given input parameters of the central engine. We then reverse the problem and develop a ``top-down'' approach in section 4, aiming at diagnosing the central engine parameters based on the observational data of the thermal emission component as well as its relative brightness with respect to the non-thermal component.  In section 5, we apply this method to GRB 110721A and derive its central engine parameters as well as their temporal evolution. Throughout the paper, the convention $Q = 10^nQ_n$ is adopted for cgs units.

\section{Hybrid system and its dynamical evolution}\label{sec:dynamics}

The aceleration of a GRB jet may be proceeded with two mechanisms: thermally driven or magnetically driven. The former is relevant for a hot fireball, which proceeds very rapidly; whereas the latter is relevant for a Poynting flux dominated outflow, which proceeds relatively more slowly. In realistic central engine models invoking either a hyper-accreting black hole or a rapidly spinning magnetar, the central engine very likely carries two components: a ``hot'' component due to neutrino heating from the accretion disk or the proto neutron star, and a ``cold'' component related to a Poynting flux launched from the black hole or the neutron star \citep[e.g.][]{lei13,metzger11}. The central engine may be characterized by a parameter
\begin{equation}
 \mu_0 = \frac{L_w}{\dot M c^2} = \frac{L_{h,0}  + L_{c,0}}{\dot M c^2} = \eta (1 + \sigma_0),
\end{equation}
which defines the total energy per baryon at the central engine, where $L_{h,0} = \eta \dot Mc^2$, $L_{c,0} = L_{\rm P,0}$, and $L_w$ are the luminosities of the hot component, cold (Poynting flux) component, and the entire wind, respectively. The dimentionless entropy $\eta$ defines average total energy (rest mass energy plus thermal energy) per baryon in the hot component, and the magnetization parameter $\sigma_0$ is defined as\footnote{The traditional definition of $\sigma$ is the ratio between the Poynting flux and the kinetic flux of matter. For the hybrid system studied this paper, it is more convenient to also include internal energy in the matter flux in the definition of a more generalized $\sigma$.}
\begin{equation}
 \sigma_0 \equiv \frac{L_c}{L_h} = \frac{L_{\rm P}}{\eta \dot M c^2}.
\end{equation}
For a variable central engine, all the parameters are a function of $t$. For each slice of wind materials (launched within a short time interval), all the parameters are a function of $r$, the radius from the central engine. 

Regardless of how the parameters evolve with the radius $r$, at any distance, one may define a parameter \citep{zhang14}
\begin{equation}
 \mu (r) = \Gamma (r) \Theta (r) (1 + \sigma (r)),
\end{equation}
where $\Gamma$ is the bulk Lorentz factor, $\Theta$ is the total co-moving energy per baryon ($\Theta-1$ is the thermal energy), and $\sigma$ is the ratio between comoving cold (magnetic) and hot (matter) energy densities. All the parameters are a function of $r$, and should evolve with $r$. If no additional baryon loading occurs and if there is no energy loss in the system, one has a conserved $\mu$ value
\begin{equation}
 \mu = \Gamma\Theta(1+\sigma)= \eta(1+\sigma_0) = \mu_0  = {\rm const}.
\end{equation}
In reality, leakage of radiation energy (the GRB emission itself) is inevitable, so that one should have $\mu < \mu_0$.

The dynamical evolution (the evolution of $\Gamma(r)$ and $\sigma(r)$) of a hybrid system is not studied in detail.  The jet dynamics in the two extreme cases, however, are well studied. 

For $\eta \gg 1$ and $\sigma_0 \ll 1$ (a pure fireball), $\g$ would firstly increase linearly with $r$ until reaching the maximum Lorentz factor defined by $\eta$ (or a characteristic value $\eta_c$ if $\eta > \eta_c$, \citealt{meszaros00}), then ``coasts'' at the maximum value until the IS radius, where fast shells catch up with slow shells and dissipate kinetic energy through shocks. After the IS phase, the average Lorentz factor of the flow decreases, since a significant amount of energy is lost in the form of radiation. The ejecta keeps coasting (probably with minor decrease of Lorentz factor due to residue ISs, e.g. \citealt{liwaxman08}) until reaching the deceleration radius beyond which the inertia from the circumburst medium is large enough and the Lorentz factor of the blastwave decreases as a power law with radius (e.g. $\g \propto r^{-3/2}$ for a constant density medium).

For $\eta \sim 1$ and $\sigma_0 \gg 1$ (a Poynting flux dominated outflow), the jet dynamics is more complicated. Generally, the flow can be quickly accelerated to a ``magneto-sonic point'', beyond which the jet front loses causal contact from the engine. For a cold Poynting flux dominated outflow, the fast magnetosonic speed is essentially the Alfven speed. For a high-$\sigma$ flow, the ``Alfvenic'' Lorentz factor is
\begin{equation}
\g_A=(1+\s)^{1/2}.
\end{equation} \label{gammaA}
The magnetosonic point, which we call the radius of rapid acceleration $r_{\rm ra}$ in the rest of the paper, is defined when the bulk Lorentz factor equals the ``Alfvenic'' Lorentz factor, so that \citep{li92,komissarov09,kumar15}
\begin{equation}
\g_{\rm ra}=(1+\s_0)^{1/3} \simeq \s_0^{1/3}
\end{equation} 
and 
\begin{equation}
(1+\s_{\rm ra}) = (1+\s_0)^{2/3},
\end{equation}
or $\s_{\rm ra} \simeq \s_0^{2/3}$. Here $\g_{\rm ra}=\g(r_{\rm ra})$ and $\s_{\rm ra}=\s(r_{\rm ra})$ are the Lorentz factor and $\s$ value at $r_{\rm ra}$, respectively. The acceleration law during this rapid acceleration phase may be written as a power law, $\g \propto r^{\lambda}$, with the power-law index $\lambda$ defined by the geometric configuration of the GRB jet. The value of $\lambda$ may be between 1/2 and 1 \citep[e.g.][]{komissarov09,granot11}. Since the exact value does not affect the main conclusion of this work, for simplicity, we adopt $\lambda=1$ in the rest of the discussion, so that this acceleration phase can be treated as similar to the thermally-driven case. The caveat of this assumption will be discussed in detail in section 6.

Above $r_{\rm ra}$, since $\s$ is still $\gg 1$, continued acceleration of the ejecta is still possible. However, the acceleration is slow and delicate, depending on the detailed magnetic configuration, and whether there is significant magnetic dissipation along the way. 
The most rapid acceleration would have a power law form $\g \propto r^{1/3}$, either due to continuous magnetic dissipation \citep{drenkhahn02} or via an impulsive acceleration mechanism \citep{granot11}. In general, such an process may be described by a general acceleration law $\g \propto r^{\delta}$ with $0<\delta\leq1/3$ \citep[e.g.][]{meszaros11,veres12}. Ideally, the acceleration would continue until reaching the coasting radius $r_c$ where $\g$ reaches $\s_0$. However, if $\s_0$ is large enough, the jet may start to decelerate before the maximum $\g$ is reached ($r_{\rm dec} < r_c$, e.g. \citealt{granot12}). Furthermore, due to the internal irregularity of the outflow, multiple internal collisions within the moderately high-$\s$ jet would trigger ICMART events to dissipate magnetic energy \citep{zhangyan11}. This would lead to a sudden drop of $\s$ and an abrupt increase of $\g$ in the emission region \citep[e.g.][]{zhangzhang14}. If the ejecta are individual magnetic blobs \citep[e.g.][]{yuanzhang12}, magnetic dissipation is facilitated since one collision would trigger significant reconnection activities, as verified by numerical simulations (W. Deng et al. 2015, in preparation).

When a Poynting flux propagates inside a star, the jet may be collimated by the stellar envelope. The confinement may lead to an additional magnetic acceleration once the jet exits the star \citep{tchekhovskoy09}. Such an acceleration critically depend on the collimation profile inside the stellar envelope. \cite{tchekhovskoy09} assumed $\theta(r) \propto r^{-\nu/2}$ so that the confine pressure is a power law function of radius $p \propto r^{-\alpha}$, with $\alpha = 2 (2-\nu)$. In their simulation, they specifically adopted $\nu=3/4$ so that $p \propto r^{-5/2}$. The reason to adopt this value is that it is consistent with the profile of a jet-shocked stellar envelope during the propagation of a relativistic jet inside a star \citep[e.g.][]{zhangw03}. As a result, the additional acceleration phase depicted by \cite{tchekhovskoy09} would be relevant for the very early episode of jet emission when it first breaks out the envelope. During the majority of the jet emission phase, a continuous jet launched from the central engine would pass freely through the already opened envelope. We speculate that the extra confinement effect discussed by \cite{tchekhovskoy09} would not play an essential role, and therefore ignore this effect in the treatment below. Nonetheless, we caution that the results would be affected if this collimation factor turns out to be important.

For a more complicated hybrid jet system, we make the assumption that acceleration proceeds first thermally and then magnetically \citep[e.g.][]{meszaros97,vlahakis03}. Since thermal acceleration proceeds linearly, and the early magnetic acceleration below the magneto-sonic point also proceeds rapidly, we approximately assume that the ejecta first gets accelerated with $\g \propto r$ until reaching a more generally defined \emph{rapid acceleration} radius $r_{\rm ra}$ defined by the larger one of the thermal coasting radius and the magneto-sonic point. Even though magnetic acceleration may deviate from the linear law below $r_{\rm ra}$, the mix with thermal acceleration would make the acceleration law in this phase very close to linear.

There are two situations.
If $\eta > (1+\s_0)^{1/2}$, after the linear acceleration phase of a fireball, the Lorentz factor of the magnetized outflow already exceeds its Alfven Lorentz factor, so that no rapid acceleration can be proceeded magnetically. One therefore has, for $\eta > (1+\s_0)^{1/2}$,
\begin{eqnarray}
\g_{\rm ra} & = & \frac{\eta}{\Theta_{\rm ra}},  \nonumber \\
1+\s_{\rm ra} & = & 1+\s_0.
\end{eqnarray}
Notice that $\s$ essentially does not decrease during this phase, but the matter portion of the luminosity changes from the thermal form to the kinetic form. 

In the opposite regime, i.e. $\eta< (1+\s_0)^{1/2}$, after the thermal acceleration phase, the outflow still moves with a Lorentz factor smaller than $\g_A$ (Eq.(\ref{gammaA})), so that it can still undergo rapid acceleration until $\g_{\rm ra} = \g_A$ is satisfied. One therefore has, for $\eta < (1+\s_0)^{1/2}$,
\begin{eqnarray}
\g_{\rm ra} & = & \left[\frac{\eta}{\Theta_{\rm ra}} (1+\sigma_0)\right]^{1/3},  \nonumber \\
1+\s_{\rm ra} & = & \left[\frac{\eta}{\Theta_{\rm ra}} (1+\sigma_0)\right]^{2/3}.
\end{eqnarray}
Putting these together, one can generally define
\begin{equation}
\g_{\rm ra}={\rm max}\left(\frac{\eta}{\Theta_{\rm ra}},\left[\frac{\eta}{\Theta_{\rm ra}} (1+\sigma_0)\right]^{1/3}\right),
\label{Gamma_ra}
\end{equation} 
and 
\begin{equation}
1+\s_{\rm ra}={\rm min}\left(1+\s_0,\left[\frac{\eta}{\Theta_{\rm ra}} (1+\sigma_0)\right]^{2/3} \right).
\end{equation} 
Here $\Theta_{\rm ra} \sim 1$ is the total co-moving energy per baryon at $r_{\rm ra}$. 

Beyond $r_{\rm ra}$, the jet would undergo a relatively slow acceleration with $\g \propto r^{\delta}$ until reaching a \emph{coasting radius} $r_{c}$. If one ignores radiation energy loss, the coasting Lorentz factor would be
\begin{equation}
\g_{c}=\frac{\eta(1+\s_0)}{\Theta_c(1+\s_c)} \simeq \eta(1+\s_0), 
\label{rc}
\end{equation}
since $\Theta_c \sim 1$, and $\s_c \ll 1$. Here $\g_c$, $\s_c$, and $\Theta_c$ are the Lorentz factor, magnetization parameter, and comoving energy per baryon at $r_c$, respectively. 

%Here $\xi < 1$ denotes the reducing factor due to radiation energy loss before reaching the coasting radius. For a highly magnetized flow, it is possible that abrupt, significant energy dissipation (e.g. ICMART) may occur before coasting is reached. Even without such abrupt dissipation events, energy loss can occur if magnetic dissipation continues to proceed steadily (see below for more discussion).

In summary, if one ignores deceleration and energy loss,  the $\Gamma$ evolution for a hybrid system may be approximated as
\begin{eqnarray}
\label{eq:gammanodis} 
\Gamma(r) = \left\{ \begin{array}{ll} \frac{r}{r_0}, & r_0<r<r_{\rm ra};\\
\g_{\rm ra}\left(\frac{r}{r_{\rm ra}}\right)^{\delta}, &
 r_{\rm ra}<r<r_{\rm c}; \\
\g_{\rm c}, &
r>r_{\rm c}, \\
\end{array} \right.
\end{eqnarray}
where $r_0$ is the radius of the jet base (size of the central engine), $r_{\rm ra}=\g_{\rm ra} r_0$, $\g_{\rm ra}$ follows eq.(\ref{Gamma_ra}), 
\begin{equation}
r_{c}=r_{\rm ra}\left(\frac{\g_{c}}{\g_{\rm ra}}\right)^{1/\delta},
\end{equation} 
and $\g_{\rm c}$ follows Eq.(\ref{rc}).

\begin{figure}[h!!!]
\begin{minipage}[b]{0.5\textwidth}
\centering \psfig{file=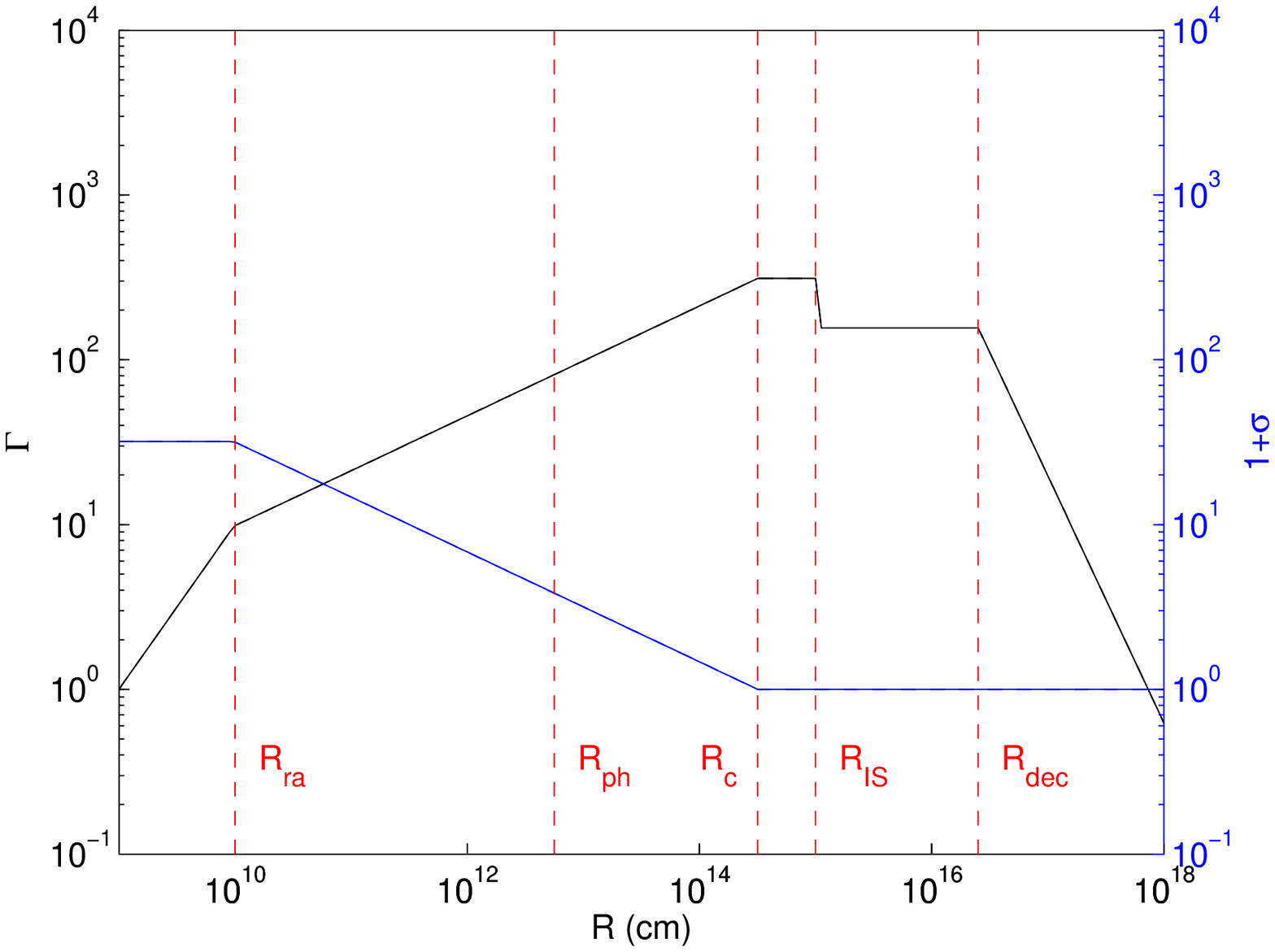,width=2.5in}
\end{minipage}\\
\begin{minipage}[b]{0.5\textwidth}
\centering \psfig{file=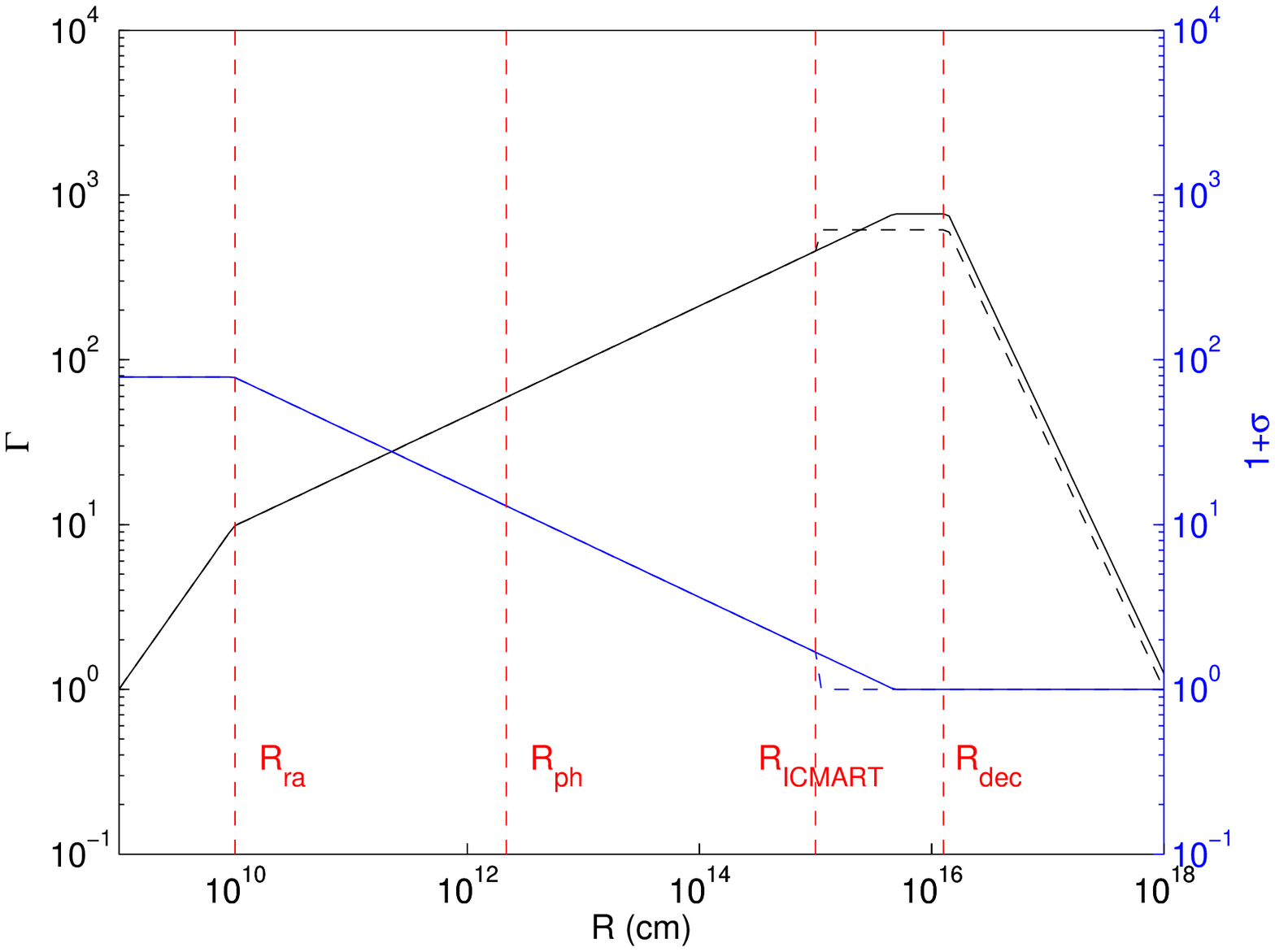,width=2.5in}
\end{minipage}\\
\begin{minipage}[b]{0.5\textwidth}
\centering \psfig{file=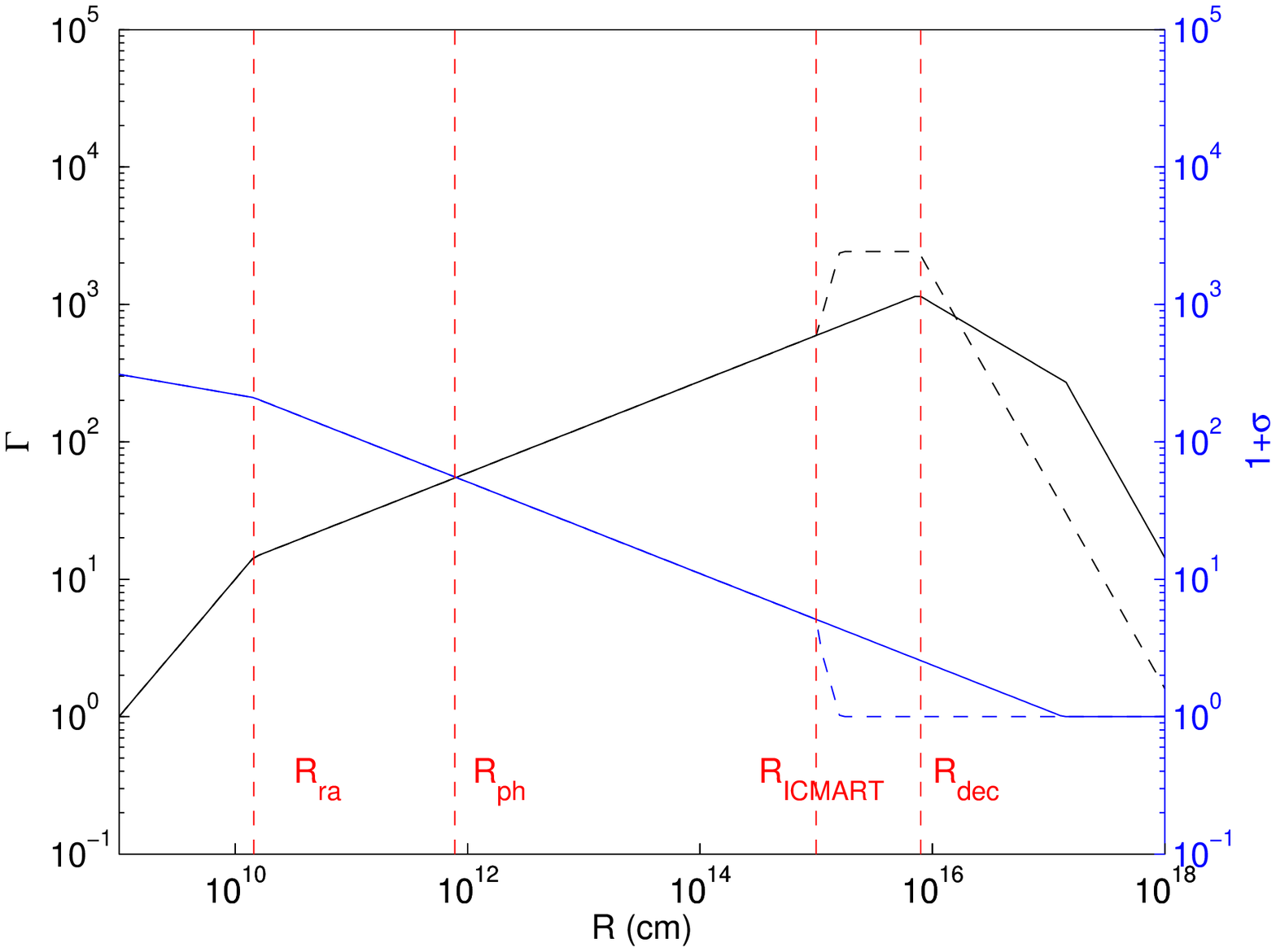,width=2.5in}
\end{minipage}%
      \caption{Examples of GRB jet dynamics. Black lines are for $\g$ evolution and blue lines are for $\s$ evolution. Vertical dashed lines denote some characteristic radii: rapid acceleration radius $r_{\rm ra}$, photosphere radius $r_{\rm ph}$, coasting radius $r_c$, internal shock radius $r_{_{\rm IS}}$, ICMART radius $r_{_{\rm ICMART}}$, and deceleration radius $r_{\rm dec}$. Following parameters are adopted: $L_w=10^{52} {\rm erg~s^{-1}}$, and $r_0=10^9 {\rm cm}$.  Different panels correspond to different combinations of $\eta$ and $\s_0$. Top panel: $\eta=10$ and $\s_0=30$. Internal shocks can form, which dissipate energy and reduce the total kinetic energy in the system; Middle panel: $\eta=10$ and $\s_0=80$; Bottom panel, $\eta=10$ and $\s_0=300$. In both cases, $\s$ is above unity at $10^{15}$ cm, suggesting that ICMART events may be the main mechanism to dissipate magnetic energy and power non-thermal radiation. The dashed lines denote the consequences of ICMART events: an abrupt reduction of $\s$ and a sudden acceleration of the system \citep{zhangyan11}.}
%Dashed lines is used when ICMART events happen. Here we adopt a set of typical parameter values: $L_w=10^{52} {\rm erg~s^{-1}}$, $r_0=10^9 {\rm cm}$, and $z=1$. }
           \label{fig:dynamics}
            \end{figure}

In reality, one has to consider jet deceleration at a radius $r_{\rm dec}$, as well as possible internal energy dissipation and radiation loss at the IS radius, $r_{\rm IS}$ (if dissipation occurs during the coasting phase), or at the ICMART radius, $r_{\rm ICMART}$ (if dissipation happens during the slow acceleration phase where $\s$ is still $> 1$). Which situation occurs depends on the initial condition $(\eta, \s_0)$. In Figure \ref{fig:dynamics}, we present the evolution of $\g$ and $\s$ with respect to the radius $r$ for different input parameters.  We assume the ICMART radius ($r_{\rm ICMART}$) and IS radius  ($r_{\rm IS}$) are both at $10^{15} {\rm cm}$, with radius defined as $r_{\rm IS}$ if it is in the coasting phase, but as $r_{\rm ICMART}$ if it is in the slow acceleration phase. If $\s(r_{15}) > 1$, ICMART events would occur, which would increase $\g$ and reduce $\s$ dramatically \citep{zhangyan11}.
The deceleration radius $r_{\rm dec}$ here is defined as the radius where the total energy of the swept-up matter from the interstellar medium (ISM) is half of the kinetic energy of the jet. We assume $E_{\rm K}=10^{52}~{\rm erg}$ and the number density of ISM $n=1~{\rm cm^{-3}}$ to calculate $r_{\rm dec}$. The results suggest that unless $\s_0$ is relatively small (say, below 80 for $\eta=10$), the energy dissipation region (non-thermal emission region) is generally in the slow acceleration phase where $\s > 1$, so that ICMART rather than IS would be the main mechanism to power the observed non-thermal emission from GRBs. This point is also obvious in view that for typical parameters, the value of the derived $r_c$ is typically $> 10^{15}$ cm.

\section{Photosphere emission from a hybrid jet: the bottom-up approach}\label{sec:bottom-up}

The photosphere radius, $r_{\rm ph}$, is defined by the condition that the photon optical depth for Thomson scattering drops below unity, so that photons previously trapped in the fireball can escape. In the lab frame, this condition is written as
\begin{eqnarray}
\tau =\int_{r_{\rm ph}}^{\infty} n_e\sigma_{\rm T}ds= 1,
\label{eq:tau}
\end{eqnarray}
where $\sigma_{\rm T}$ is the Thomson cross section, the lab frame electron number density can be written as
\begin{eqnarray}
n_e=\frac{L_{w} \mathcal{V}}{4\pi r^2m_p c^3 \eta (1+\s_0)},
\label{eq:ne}
\end{eqnarray}
and 
\begin{equation}
 ds = (1-\beta \cos\theta) dr/\cos\theta
\end{equation}
is the spatial increment in the outflow wind in the lab frame (e.g. Eq.(23) of \citealt{deng14}), $\theta$ is the angle from line-of-sight, $m_p$ is proton mass, $c$ is speed of light, and $\mathcal{V}$ is the lepton-to-baryon number ratio. We assume that $\mathcal{V} \ll m_p/m_e$ is satisfied, so that the inertia of the leptons is negligible. Consider the line of sight direction ($\theta=0$), one has $ds = (1-\beta) dr \simeq dr/(2\g^2)$. For $\g$ evolution as shown in Equation (\ref{eq:gammanodis}), the line-of-sight photosphere radius can be derived as
\begin{eqnarray}
\label{eq:rph} r_{\rm ph} = \left\{ \begin{array}{ll} \left(\frac{L_w \mathcal{V} \sigma_{\rm T}r_0^{2} }{8\pi m_p c^3 \eta(1+\sigma_0) }\right)^{1/3}, & r_0<r_{\rm ph}<r_{\rm ra};\\
\left(\frac{L_w \mathcal{V} \sigma_{\rm T}r_{\rm ra}^{2\delta}}{8\pi m_p c^3 \g_{\rm ra}^2 \eta (1+\sigma_0)}\right)^{1/(2\delta +1)}, &
r_{\rm ra}<r_{\rm ph}<r_{\rm c};  \\
\frac{L_w \mathcal{V} \sigma_{\rm T}}{8\pi m_p c^3 \g_{\rm c}^2\eta (1+\sigma_0)}, &
r_{\rm ph}>r_{\rm c}. \\
\end{array} \right.
\end{eqnarray}

As shown in \cite{pe'er07}, for quasi-thermal photosphere emission as expected in a non-dissipative photosphere, the observed temperature and thermal flux can be derived as 
\begin{eqnarray}
T_{\rm ob}& = & C_1\g_{\rm ph} T'_{\rm ph} /(1+z),\\
F_{\rm BB} & = & \mathcal{R}^2\sigma_{\rm SB} T_{\rm ob}^4,
\end{eqnarray}
where 
\begin{eqnarray}
\mathcal{R}=C_2~\frac{(1+z)^2}{d_L}\frac{r_{\rm ph}}{\g_{\rm ph}},
\end{eqnarray}
$z$ is the redshift, $d_L$ is the luminosity distance, $\sigma_{\rm SB}$ is the Stefan-Boltzmann constant, $T'_{\rm ph}$ is the comoving temperature at the photosphere radius $r_{\rm ph}$, and $C_1 \simeq 1.48$ and $C_2 \simeq 1.06$ are the factors derived from detailed numerical integration of angle- and distance-dependent photosphere emission.

The photosphere properties depends on whether significant magnetic energy dissipation happens below the photosphere radius $r_{\rm ph}$. This is an open question, and no firm conclusion has been drawn from the first principles. In the following, we discuss both scenarios.

\subsection{The case of no magnetic dissipation}\label{sec:no-mag-bottom}

This scenario assumes that no magnetic field reconnection occurs below the photosphere, so that no magnetic energy is directly converted to particle energy and heat. Magnetic acceleration in any case proceeds, so that some magnetic energy is converted to the kinetic energy of the outflow. Such a scenario may be relevant to helical jets or self-sustained magnetic bubbles \citep[e.g.][]{spruit01,uzdensky06,yuanzhang12}. This scenario predicts a quasi-thermal photosphere emission component, which is consistent with the data of several Fermi GRBs \citep[e.g.][]{ryde10,zhang11,guiriec11,axelsson12,guiriec13}.

Without magnetic heating, the thermal energy undergoes adiabatic cooling, with $r^2e^{3/4}\g={\rm const}$ \citep[e.g.][]{piran93}. Noticing $e \propto {T'}^4$ and the dynamical evolution eq.(\ref{eq:gammanodis}), one can derive the comoving temperature at the photosphere radius $r_{\rm ph}$ as
 \begin{eqnarray}
\label{eq:Tph1} 
T'_{\rm ph} = \left\{ \begin{array}{ll} T_0\left(\frac{r_{\rm ph}}{r_0}\right)^{-1}, & r_0<r_{\rm ph}<r_{\rm ra};\\
T_0\left(\frac{r_{\rm ra}}{r_0}\right)^{-1}\left(\frac{r_{\rm ph}}{r_{\rm ra}}\right)^{-(2+\delta)/3}, &
r_{\rm ra}<r_{\rm ph}<r_{\rm c};  \\
T_0\left(\frac{r_{\rm ra}}{r_0}\right)^{-1}\left(\frac{r_{\rm c}}{r_{\rm ra}}\right)^{-(2+\delta)/3}\left(\frac{r_{\rm ph}}{r_{\rm c}}\right)^{-2/3}, &
r_{\rm ph}>r_{\rm c}. \\
\end{array} \right.
\end{eqnarray}
Here 
\begin{equation}
T_0 \simeq \left(\frac{L_w}{4\pi r_0^2 a c (1+\sigma_0)}\right)^{1/4}
\end{equation} 
is the temperature at $r_0$, $a=7.56 \times 10^{-15} {\rm erg~cm^{-3}~K^{-4}}$ is radiation density constant. Given the central engine parameters $L_w$, $r_0$, $\eta$ and $\s_0$, we can derive all the relevant photosphere properties with equations from (\ref{eq:gammanodis}) to (\ref{eq:Tph1}), as long as the slow magnetic acceleration index $\delta$ is determined.  The largest $\delta$ is 1/3, which is achievable for an impulsive, non-dissipative magnetic shell \citep{granot11}. The general expressions for an arbitrary $\delta$ are presented in the Appendix. In the following, we present the results for $\delta=1/3$. The implications of an arbitrary $\delta$ values are discussed in detail in Section \ref{sec:discussion}. Also in the following analytical formulae, we have adopted $\Theta_{\rm ra}=\Theta_{c}=1$ (cold flow) and $\s_c=0$ as a reasonable approximation.

For different central engine parameters, $\g_{\rm ra}$ can have two possible values: $\eta$ or $[\eta (1+\s_0)]^{1/3}$. For each case, the photosphere radius $r_{\rm ph}$ can be in three different regimes separated by $r_{\rm ra}$ and $r_c$. So altogether we can define six different regimes: (I) $\eta>(1+\s_0)^{1/2}$ and $r_{\rm ph}<r_{\rm ra}$; (II) $\eta>(1+\s_0)^{1/2}$ and $r_{\rm ra}<r_{\rm ph}<r_{c}$; (III) $\eta>(1+\s_0)^{1/2}$ and $r_{\rm ph}>r_{c}$;  (IV) $\eta<(1+\s_0)^{1/2}$ and $r_{\rm ph}<r_{\rm ra}$; (V) $\eta<(1+\s_0)^{1/2}$ and $r_{\rm ra}<r_{\rm ph}<r_{c}$; (VI) $\eta<(1+\s_0)^{1/2}$ and $r_{\rm ph}>r_{c}$. The six regimes also apply for the case with significant magnetic dissipation below the photosphere (see below). In Table 1, we list the criteria of all 12 regimes (for the cases of both without and with magnetic dissipations) based on the central engine properties. In the following, we derive relevant parameters in each regime, including $r_{\rm ra}$ and $r_{c}$ (which are useful to write down the $\g$ evolution of the system), along with the photosphere properties, i.e. $r_{\rm ph}$, $\g_{\rm ph}$, $(1+\sigma_{\rm ph})$, $kT_{\rm ob}$, and $F_{\rm BB}$:
% (the presented result of $T_{\rm ob}$ is actually for $k T_{\rm  ob}$ in unit of kev):

Regime I:

\begin{eqnarray}
&&r_{\rm ra}=1.0\times 10^{11}~{\rm cm}~r_{0,9}\eta_2 ,\nonumber\\
&&r_{\rm c}=1.0\times 10^{17}~{\rm cm}~r_{0,9}\eta_2(1+\s_0)_2^{3} ,\nonumber\\
&&r_{\rm ph}=8.34\times 10^{10}~{\rm cm}~L_{w,52}^{1/3}r_{0,9}^{2/3}\eta_2^{-1/3}(1+\s_0)_2^{-1/3} ,\nonumber\\
&&\g_{\rm ph}=83.4L_{w,52}^{1/3}r_{0,9}^{-1/3}\eta_2^{-1/3}(1+\s_0)_2^{-1/3},\nonumber\\
&&1+\s_{\rm ph}=100(1+\s_0)_2,\nonumber\\
&&k T_{\rm ob}=56.1~{\rm keV}~(1+z)^{-1}L_{w,52}^{1/4}r_{0,9}^{-1/2}(1+\s_0)_2^{-1/4},\nonumber\\
&&F_{\rm BB}=1.07\times 10^{-7}~{\rm erg~s^{-1}cm^{-2}}~L_{w,52}(1+\s_0)_2^{-1}d_{L,28}^{-2}.\nonumber\\
\end{eqnarray}

Regime II:

\begin{eqnarray}
&&r_{\rm ra}=1.0\times 10^{11}~{\rm cm}~r_{0,9}\eta_2 ,\nonumber\\
&&r_{\rm c}=1.0\times 10^{17}~{\rm cm}~r_{0,9}\eta_2(1+\s_0)_2^{3},\nonumber\\
&&r_{\rm ph}=7.22\times 10^{10}~{\rm cm}~L_{w,52}^{3/5}r_{0,9}^{2/5}\eta_2^{-7/5}(1+\s_0)_2^{-3/5},\nonumber\\
&&\g_{\rm ph}=89.7L_{w,52}^{1/5}r_{0,9}^{-1/5}\eta_2^{1/5}(1+\s_0)_2^{-1/5},\nonumber\\
&&1+\s_{\rm ph}=17.7L_{w,52}^{-1/5}r_{0,9}^{1/5}\eta_2^{4/5}(1+\s_0)_2^{6/5},\nonumber\\
&&k T_{\rm ob}=64.8~{\rm keV}~(1+z)^{-1}L_{w,52}^{-1/60}r_{0,9}^{-7/30}\eta_2^{16/15}(1+\s_0)_2^{1/60},\nonumber\\
&&F_{\rm BB}=1.24\times 10^{-7}~{\rm erg~s^{-1}cm^{-2}}~L_{w,52}^{11/15}r_{0,9}^{4/15}\eta_2^{16/15}(1+\s_0)_2^{-11/15}d_{L,28}^{-2}.\nonumber\\
\end{eqnarray}

Regime III:

\begin{eqnarray}
&&r_{\rm ra}=1.0\times 10^{11}~{\rm cm}~r_{0,9}\eta_2,\nonumber\\
&&r_{\rm c}=1.0\times 10^{17}~{\rm cm}~r_{0,9}\eta_2(1+\s_0)_2^{3},\nonumber\\
&&r_{\rm ph}=5.81\times 10^{12}~{\rm cm}~L_{w,52}\eta_1^{-3}(1+\s_0)_1^{-3},\nonumber\\
&&\g_{\rm ph}=100\eta_1(1+\s_0)_1,\nonumber\\
&&1+\s_{\rm ph} \simeq 1,\nonumber\\
&&k T_{\rm ob}=6.65~{\rm keV}~(1+z)^{-1}L_{w,52}^{-5/12}r_{0,9}^{1/6}\eta_1^{8/3}(1+\s_0)_1^{29/12},\nonumber\\
&&F_{\rm BB}=7.15\times 10^{-8}~{\rm erg~s^{-1}cm^{-2}}~L_{w,52}^{1/3}r_{0,9}^{2/3}\eta_1^{8/3}(1+\s_0)_1^{5/3}d_{L,28}^{-2}.\nonumber\\
\end{eqnarray}

Regime IV:

\begin{eqnarray}\\
&&r_{\rm ra}=2.15\times 10^{10}~{\rm cm}~r_{0,9}\eta_2^{1/3}(1+\s_0)_2^{1/3},\nonumber\\
&&r_{\rm c}=2.15\times 10^{18}~{\rm cm}~r_{0,9}\eta_2^{7/3}(1+\s_0)_2^{7/3},\nonumber\\
&&r_{\rm ph}=8.34\times 10^{10}~{\rm cm}~L_{w,52}^{1/3}r_{0,9}^{2/3}\eta_2^{-1/3}(1+\s_0)_2^{-1/3},\nonumber\\
&&\g_{\rm ph}=83.4L_{w,52}^{1/3}r_{0,9}^{-1/3}\eta_2^{-1/3}(1+\s_0)_2^{-1/3},\nonumber\\
&&1+\s_{\rm ph}=5.56L_{w,52}^{-1/3}r_{0,9}^{1/3}\eta_2^{4/3}(1+\s_0)_2^{4/3},\nonumber\\
&&k T_{\rm ob}=56.1~{\rm keV}~(1+z)^{-1}L_{w,52}^{1/4}r_{0,9}^{-1/2}(1+\s_0)_2^{-1/4},\nonumber\\
&&F_{\rm BB}=1.07\times 10^{-7}~{\rm erg~s^{-1}cm^{-2}}~L_{w,52}(1+\s_0)_2^{-1}d_{L,28}^{-2}.\nonumber\\
\end{eqnarray}

Regime V:

\begin{eqnarray}
&&r_{\rm ra}=2.15\times 10^{10}~{\rm cm}~r_{0,9}\eta_2^{1/3}(1+\s_0)_2^{1/3},\nonumber\\
&&r_{\rm c}=2.15\times 10^{18}~{\rm cm}~r_{0,9}\eta_2^{7/3}(1+\s_0)_2^{7/3},\nonumber\\
&&r_{\rm ph}=2.46\times 10^{11}~{\rm cm}~L_{w,52}^{3/5}r_{0,9}^{2/5}\eta_2^{-13/15}(1+\s_0)_2^{-13/15},\nonumber\\
&&\g_{\rm ph}=48.5L_{w,52}^{1/5}r_{0,9}^{-1/5}\eta_2^{-1/15}(1+\s_0)_2^{-1/15},\nonumber\\
&&1+\s_{\rm ph}=17.7L_{w,52}^{-1/5}r_{0,9}^{1/5}\eta_2^{16/15}(1+\s_0)_2^{16/15},\nonumber\\
&&k T_{\rm ob}=19.0~{\rm keV}~(1+z)^{-1}L_{w,52}^{-1/60}r_{0,9}^{-7/30}\eta_2^{8/15}(1+\s_0)_2^{17/60},\nonumber\\
&&F_{\rm BB}=3.63\times 10^{-8}~{\rm erg~s^{-1}cm^{-2}}~L_{w,52}^{11/15}r_{0,9}^{4/15}\eta_2^{8/15}(1+\s_0)_2^{-7/15}d_{L,28}^{-2}.\nonumber\\
\end{eqnarray}

Regime VI:

\begin{eqnarray}
&&r_{\rm ra}=2.15\times 10^{10}~{\rm cm}~r_{0,9}\eta_2^{1/3}(1+\s_0)_2^{1/3},\nonumber\\
&&r_{\rm c}=2.15\times 10^{18}~{\rm cm}~r_{0,9}\eta_2^{7/3}(1+\s_0)_2^{7/3},\nonumber\\
&&r_{\rm ph}=5.81\times 10^{12}~{\rm cm}~L_{w,52}\eta_1^{-3}(1+\s_0)_1^{-3},\nonumber\\
&&\g_{\rm ph}=100\eta_1(1+\s_0)_1,\nonumber\\
&&1+\s_{\rm ph} \simeq 1,\nonumber\\
&&k T_{\rm ob}=6.65~{\rm keV}~(1+z)^{-1}L_{w,52}^{-5/12}r_{0,9}^{1/6}\eta_1^{8/3}(1+\s_0)_1^{29/12},\nonumber\\
&&F_{\rm BB}=7.15\times 10^{-8}~{\rm erg~s^{-1}cm^{-2}}~L_{w,52}^{1/3}r_{0,9}^{2/3}\eta_1^{8/3}(1+\s_0)_1^{5/3}d_{L,28}^{-2}.\nonumber\\
\end{eqnarray}

To better present our results, we show the contour plots of $k T_{\rm ob}$ and $F_{\rm BB}$ in the ($\eta,1+\s_0$) plane in Figure \ref{fig:contour1}.

\subsection{The case of magnetic dissipation}\label{sec:mag-bottom}

It has been speculated that significant magnetic dissipation may occur during the propagation of the jet below the photosphere. Such a magnetically dissipative photosphere \citep[e.g.][]{thompson94,rees05,giannios08,meszaros11,veres12} would lead to enhancement of photosphere emission. On the other hand, the photosphere emission behavior is determined by the physical conditions far below the photosphere radius, where complete thermalization is not necessarily achieved without efficient creation of additional photons \citep{pe'er06,giannios06,beloborodov10,beloborodov13,levinson12,vurm13,begue14}. Specifically, for a dissipative photosphere due to magnetic dissipation, recent studies show that the photosphere emission could have a non-thermal appearance with a spectral peak ($E_p$) varying from 1 MeV up to a maximum value of about 20 MeV, depending on magnetization fraction $\sigma_0$ \citep{beloborodov13,begue14}. Nonetheless, an {\em effective} temperature can be derived, which would be the temperature if the emission is fully thermalized \citep[e.g.][]{eichler00,thompson07}. Practically, it would serve as an estimate of the {\em lower limit} of $E_p$ of a magnetically dissipative photosphere emission.
% For a given radiative luminosity, the flux level and the minimum possible temperature of the photosphere could be estimated by setting the blackbody limit reached upon complete thermalization \citep{eichler00,thompson07}. 
In this section, we quantify the photosphere emission properties under the assumption of significant magnetic dissipation.

With magnetic dissipation, the adiabatic relation no longer applies. One needs to introduce another conservation relation. 
One natural assumption is that the magnetic energy is converted into both thermal energy and kinetic energy of the bulk motion with fixed proportions (e.g. $1:1$, but the exact proportions do not matter to define the temperature evolution of the system). With this assumption, in the lab frame and after the initial thermal acceleration phase, the internal energy should be proportional to $\g$, so that in the comoving frame energy per baryon, $\Theta(r)$, should remain constant. This is translated to $r^2 e \g = {\rm const}$ (noticing that the co-moving size increases with $\g$). In this case, the evolution of comoving temperature should be revised as 
 \begin{eqnarray}
\label{eq:rph} 
T'_{\rm ph} = \left\{ \begin{array}{ll} T_0\left(\frac{r_{\rm ph}}{r_0}\right)^{-1}, & r_0<r_{\rm ph}<r_{\rm ra};\\
T_0\left(\frac{r_{\rm ra}}{r_0}\right)^{-1}\left(\frac{r_{\rm ph}}{r_{\rm ra}}\right)^{-(2+\delta)/4}, &
r_{\rm ra}<r_{\rm ph}<r_{\rm c};  \\
T_0\left(\frac{r_{\rm ra}}{r_0}\right)^{-1}\left(\frac{r_{\rm c}}{r_{\rm ra}}\right)^{-(2+\delta)/4}\left(\frac{r_{\rm ph}}{r_{\rm c}}\right)^{-2/3}, &
r_{\rm ph}>r_{\rm c}. \\
\end{array} \right.
\end{eqnarray}
Two remarks need to be made here. First, at $r_{ph} < r_{ra}$ (the first segment), there should be a segment in which $T' \propto r^{-3/4}$ is satisfied. This is relevant for $\eta < (1+\s_0)^{1/2}$ but at $r>r_0 \eta$ (i.e. thermal acceleration is over and the flow is under rapid magnetic acceleration). The deviation from the approximate $T' \propto r^{-1}$ would be significant if $\eta \ll (1+\sigma)^{1/2}$. However, in reality, a central engine always has a reasonably ``hot'' component so that $\eta \gg 1$. Also the introduction of this additional regime would not change the results substantially. We therefore do not get into the complications of introducing these trivial regimes. Second, at $r>r_c$, one has $\s < 1$. Heating due to magnetic dissipation becomes insignificant. One goes back to the scaling for an adiabatic outflow.

Using Eq.(\ref{eq:rph}), we can similarly derive the relevant parameters for the six regimes in the magnetic dissipation case. Here $T_{\rm BB}^{\rm eff}$ is the effective temperature of the photosphere emission, so that $k T_{\rm BB}^{\rm eff}$ represents the lower limit of $E_p$ of the dissipative photosphere emission. The corresponding effective blackbody flux is denoted as $F_{\rm ph}$, which represents the $\nu F_\nu$ flux level for the photosphere emission. This flux level does not depend on the unknown $E_p$ of the dissipative photosphere emission.

Regime I:

\begin{eqnarray}
&&r_{\rm ra}=1.0\times 10^{11}~{\rm cm}~r_{0,9}\eta_2,\nonumber\\
&&r_{\rm c}=1.0\times 10^{17}~{\rm cm}~r_{0,9}\eta_2(1+\s_0)_2^{3},\nonumber\\
&&r_{\rm ph}=8.34\times 10^{10}~{\rm cm}~L_{w,52}^{1/3}r_{0,9}^{2/3}\eta_2^{-1/3}(1+\s_0)_2^{-1/3},\nonumber\\
&&\g_{\rm ph}=83.4L_{w,52}^{1/3}r_{0,9}^{-1/3}\eta_2^{-1/3}(1+\s_0)_2^{-1/3},\nonumber\\
&&1+\s_{\rm ph}=100(1+\s_0)_2,\nonumber\\
&&k T_{\rm BB}^{\rm eff}=56.1~{\rm keV}~(1+z)^{-1}L_{w,52}^{1/4}r_{0,9}^{-1/2}(1+\s_0)_2^{-1/4},\nonumber\\
&&F_{\rm ph}=1.07\times 10^{-7}~{\rm erg~s^{-1}cm^{-2}}~L_{w,52}(1+\s_0)_2^{-1}d_{L,28}^{-2}.\nonumber\\
\end{eqnarray}

Regime II:

\begin{eqnarray}
&&r_{\rm ra}=1.0\times 10^{11}~{\rm cm}~r_{0,9}\eta_2,\nonumber\\
&&r_{\rm c}=1.0\times 10^{17}~{\rm cm}~r_{0,9}\eta_2(1+\s_0)_2^{3},\nonumber\\
&&r_{\rm ph}=7.22\times 10^{10}~{\rm cm}~L_{w,52}^{3/5}r_{0,9}^{2/5}\eta_2^{-7/5}(1+\s_0)_2^{-3/5},\nonumber\\
&&\g_{\rm ph}=89.7L_{w,52}^{1/5}r_{0,9}^{-1/5}\eta_2^{1/5}(1+\s_0)_2^{-1/5},\nonumber\\
&&1+\s_{\rm ph}=17.7L_{w,52}^{-1/5}r_{0,9}^{1/5}\eta_2^{4/5}(1+\s_0)_2^{6/5},\nonumber\\
&&k T_{\rm BB}^{\rm eff}=60.8~{\rm keV}~(1+z)^{-1}L_{w,52}^{1/10}r_{0,9}^{-7/20}\eta_2^{3/5}(1+\s_0)_2^{-1/10},\nonumber\\
&&F_{\rm ph}=9.62\times 10^{-8}~{\rm erg~s^{-1}cm^{-2}}~L_{w,52}^{6/5}r_{0,9}^{-1/5}\eta_2^{-4/5}(1+\s_0)_2^{-6/5}d_{L,28}^{-2}.\nonumber\\
\end{eqnarray}

Regime III:

\begin{eqnarray}
&&r_{\rm ra}=1.0\times 10^{11}~{\rm cm}~r_{0,9}\eta_2,\nonumber\\
&&r_{\rm c}=1.0\times 10^{17}~{\rm cm}~r_{0,9}\eta_2(1+\s_0)_2^{3},\nonumber\\
&&r_{\rm ph}=5.81\times 10^{12}~{\rm cm}~ L_{w,52}\eta_1^{-3}(1+\s_0)_1^{-3},\nonumber\\
&&\g_{\rm ph}=100 \eta_1(1+\s_0)_1,\nonumber\\
&&1+\s_{\rm ph} \simeq 1,\nonumber\\
&&k T_{\rm BB}^{\rm eff}=25.5~{\rm keV}~(1+z)^{-1}L_{w,52}^{-5/12}r_{0,9}^{1/6}\eta_1^{8/3}(1+\s_0)_1^{3},\nonumber\\
&&F_{\rm ph}=1.54\times 10^{-5}~{\rm erg~s^{-1}cm^{-2}}~ L_{w,52}^{1/3}r_{0,9}^{2/3}\eta_1^{8/3}(1+\s_0)_1^{4}d_{L,28}^{-2}.\nonumber\\
\end{eqnarray}

Regime IV:

\begin{eqnarray}
&&r_{\rm ra}=2.15\times 10^{10}~{\rm cm}~r_{0,9}\eta_2^{1/3}(1+\s_0)_2^{1/3},\nonumber\\
&&r_{\rm c}=2.15\times 10^{18}~{\rm cm}~r_{0,9}\eta_2^{7/3}(1+\s_0)_2^{7/3},\nonumber\\
&&r_{\rm ph}=8.34\times 10^{10}~{\rm cm}~L_{w,52}^{1/3}r_{0,9}^{2/3}\eta_2^{-1/3}(1+\s_0)_2^{-1/3},\nonumber\\
&&\g_{\rm ph}=83.4L_{w,52}^{1/3}r_{0,9}^{-1/3}\eta_2^{-1/3}(1+\s_0)_2^{-1/3},\nonumber\\
&&1+\s_{\rm ph}=5.56L_{w,52}^{-1/3}r_{0,9}^{1/3}\eta_2^{4/3}(1+\s_0)_2^{4/3},\nonumber\\
&&k T_{\rm BB}^{\rm eff}=56.1~{\rm keV}~(1+z)^{-1}L_{w,52}^{1/4}r_{0,9}^{-1/2}(1+\s_0)_2^{-1/4},\nonumber\\
&&F_{\rm ph}=1.07\times 10^{-7}~{\rm erg~s^{-1}cm^{-2}}~L_{w,52}(1+\s_0)_2^{-1}d_{L,28}^{-2}.\nonumber\\
\end{eqnarray}

Regime V:

\begin{eqnarray}
&&r_{\rm ra}=2.15\times 10^{10}~{\rm cm}~r_{0,9}\eta_2^{1/3}(1+\s_0)_2^{1/3},\nonumber\\
&&r_{\rm c}=2.15\times 10^{18}~{\rm cm}~r_{0,9}\eta_2^{7/3}(1+\s_0)_2^{7/3},\nonumber\\
&&r_{\rm ph}=2.46\times 10^{11}~{\rm cm}~L_{w,52}^{3/5}r_{0,9}^{2/5}\eta_2^{-13/15}(1+\s_0)_2^{-13/15},\nonumber\\
&&\g_{\rm ph}=48.5L_{w,52}^{1/5}r_{0,9}^{-1/5}\eta_2^{-1/15}(1+\s_0)_2^{-1/15},\nonumber\\
&&1+\s_{\rm ph}=17.7L_{w,52}^{-1/5}r_{0,9}^{1/5}\eta_2^{16/15}(1+\s_0)_2^{16/15},\nonumber\\
&&k T_{\rm BB}^{\rm eff}=30.5~{\rm keV}~(1+z)^{-1}L_{w,52}^{1/10}r_{0,9}^{-7/20}\eta_2^{3/10}(1+\s_0)_2^{1/20},\nonumber\\
&&F_{\rm ph}=2.42\times 10^{-7}~{\rm erg~s^{-1}cm^{-2}}~L_{w,52}^{6/5}r_{0,9}^{-1/5}\eta_2^{-2/5}(1+\s_0)_2^{-7/5}d_{L,28}^{-2}.\nonumber\\
\end{eqnarray}

Regime VI:

\begin{eqnarray}
&&r_{\rm ra}=2.15\times 10^{10}~{\rm cm}~r_{0,9}\eta_2^{1/3}(1+\s_0)_2^{1/3},\nonumber\\
&&r_{\rm c}=2.15\times 10^{18}~{\rm cm}~r_{0,9}\eta_2^{7/3}(1+\s_0)_2^{7/3},\nonumber\\
&&r_{\rm ph}=5.81\times 10^{12}~{\rm cm}~ L_{w,52}\eta_1^{-3}(1+\s_0)_1^{-3},\nonumber\\
&&1+\g_{\rm ph}=100 \eta_1(1+\s_0)_1,\nonumber\\
&&1+\s_{\rm ph} \simeq 1,\nonumber\\
&&k T_{\rm BB}^{\rm eff}=39.8~{\rm keV}~(1+z)^{-1} L_{w,52}^{-5/12}r_{0,9}^{1/6}\eta_1^{55/18}(1+\s_0)_1^{101/36},\nonumber\\
&&F_{\rm ph}=9.24\times 10^{-5}~{\rm erg~s^{-1}cm^{-2}}~ L_{w,52}^{1/3}r_{0,9}^{2/3}\eta_1^{38/9}(1+\s_0)_1^{29/9}d_{L,28}^{-2}.\nonumber\\
\end{eqnarray}

The criteria for different regimes are also presented in Table 1.

\subsection{Example spectra}

With the preparation above, one can simulate some example spectra of GRB prompt emission with the superposition of a photosphere emission component (which is thermal for a non-dissipative photosphere but could significantly deviate from the thermal form for a dissipative photosphere) and a non-thermal emission component from an optically thin region. Given a set of central engine parameters $(L_w, r_0, \eta, \sigma_0)$, the photosphere component can be quantified. A detailed simulation of the non-thermal component requires the assumptions of the explicit energy dissipation mechanism and radiation mechanism. For the purpose of this paper (which focuses on photosphere emission), we introduce the non-thermal component empirically. We take the non-thermal component as a Band function with typical observed parameters \citep{preece00,zhang11,nava11}: $E_{\rm peak}=1000~{\rm keV}$, $\alpha=-1$, and $\beta=-2.2$. The normalization of the spectrum is determined by assuming that $50\%$ of the remaining wind luminosity is converted to non-thermal emission. Such an efficiency may be relatively too high for IS model \citep[e.g.][]{kumar99,panaitescu99}, but would be reasonable for ICMART events \citep{zhangyan11}. For a large parameter space, we find $\s > 1$ at $10^{15}$ cm, so that ICMART would be a more relevant energy dissipation mechanism for non-thermal emission (e.g. Fig.\ref{fig:dynamics}). Even though no radiation mechanism is specified, we note that fast cooling synchrotron radiation at a relatively large emission radius is able to reproduce typical Band function spectra as observed \citep{uhm14}.

Some example spectra are presented in Figure \ref{fig:spectrum} for different $(\eta, \s_0)$ pairs as input parameters. For non-dissipative photosphere emission, we have assumed a blackbody spectrum. For dissipative photosphere emission, on the other hand, we only plot its flux level and the range of $E_p$ defined by $k T_{\rm BB}^{\rm eff}$ and 20 MeV as suggested by \cite{beloborodov13}. One can see that a diversity of spectrum as observed by {\em Fermi} can be reproduced for the non-dissipative photosphere case, given that $\eta$ and $\s_0$ are allowed to vary in a wide range. When $\s_0 \ll 1$ (panels (a-c), the fireball case), the photosphere emission is bright, and one would expect a bright thermal component sticking out from the Band component. This is consistent with the results of \cite{zhang09} and \cite{fan10}. In particular, when $\eta$ is large enough, the spectrum is completely dominated by the thermal component, similar to the case of GRB 090902B\footnote{The power law non-thermal emission as observed in GRB 090902B would require an additional spectral component, which may be related to inverse Compton scattering of some kind \citep{pe'er12}. This is not modeled in this paper.}. As $\eta$ reduces or $\s_0$ increases (panels (d-i)), the photosphere component becomes sub-dominant (such as GRBs 100724B and 110721A), or even completely suppressed (such as GRB 080916C and many other bursts), as observed in many {\em Fermi} GRBs. For the cases with a large $\sigma_0$, magnetic dissipation would enhance the photosphere emission as compared with the case without magnetic dissipation. However, if $E_p$ of the dissipative photosphere emission is large, the predicted spectra are inconsistent with the Fermi data \citep{begue14}.

\section{Inferring central engine parameters from the data: the top-down approach}\label{sec:top-down}

In practice, a more interesting problem is to use the observed data to diagnose the properties at the central engine. \cite{pe'er07} worked out this problem for the pure fireball model. They pointed out that if $r_{\rm ph} > r_c$, it is possible to derive $\eta$ and $r_0$ based on the observed data. Due to a degeneracy, this is impossible if $r_{\rm ph} \leq r_c$, which corresponds to a very large $\eta$ value.

In this section, we solve this problem for a more generalized hybrid outflow based on our results derived from the bottom-up approach. Since a blackbody component is predicted only in the non-dissipative photosphere model, our top-down approach only applies to the non-dissipative photosphere models. The magnetically dissipative photopshere models predict a much higher $E_p$, so they are disfavored by the data \citep{begue14}.

In general, one has three observed quantities: the observed blackbody temperature $T_{\rm ob}$, the observed blackbody flux $F_{\rm BB}$, and the observed total flux $F_{\rm ob}$ (both thermal and non-thermal included). In the pure fireball model, there are three unknowns: $L_w$, $\eta$, and $r_0$. This is why \cite{pe'er07} can solve for $\eta$ and $r_0$ from the data. In the hybrid problem, another parameter $\s_0$ is introduced, so that altogether there are four unknowns\footnote{In principle, the magnetic acceleration index $\delta$ might be another unknown parameter if it differs from 1/3. In this work, we adopt $\delta=1/3$ in the top-down approach and in the case study of GRB 110721A. The effect of a general $\delta$ value is discussed in detail at the end of Section \ref{sec:discussion}.}. It is impossible to solve all four parameters from the data. On the other hand, applying the fireball method of \cite{pe'er07} to GRB 110721A led to curious, unreasonable parameters for $\eta$ and $r_0$ as a function of $t$ \citep{iyyani13}, suggesting that there are more parameters at play at the central engine. Physically, it is more reasonable to assume an essentially constant $r_0$ throughout a burst. Indeed, \cite{ghirlanda13} found that this is the case for some GRBs that are of a fireball origin. For hybrid systems, it is reasonable to assume a constant value for $r_0$, and use the data to infer other three parameters. The results vary for different $r_0$ values. 

In the following, we derive constraints on the central engine parameters $\eta$ and $\s_0$. We define $f_\gamma = L_\gamma/L_w$, which connects the total flux $F_{\rm ob}$ to the wind luminosity $L_w$. We also define $f_{\rm th} = F_{\rm BB}/F_{\rm ob}$, which can be directly measured from the data. We then express $\eta$ and $\s_0$ in terms of the measurables $T_{\rm ob}$ and $f_{\rm th}$, along with $f_\gamma$ and $r_0$, both are taken as constants and can be estimated to a typical value (e.g. $f_\gamma = 0.5$, $r_0 = 10^8$ cm).

We also derive the radius of the photosphere $r_{\rm ph}$, the Lorentz factor $\g_{\rm ph}$, and the magnetization parameter $(1+\s_{\rm ph})$ at the photosphere. In order to check whether IS or ICMART is responsible for the non-thermal emission, we also derive the magnetization parameter at $\sim 10^{15}$ cm, $(1+\s_{15})$, based on a simple $\g \propto r^{1/3}$ scaling law.  If the derived $(1+\s_{15})$ is smaller than 1, it means that $10^{15}$ cm is already in the coasting regime, and IS should be the main mechanism for non-thermal energy dissipation \citep[e.g.][]{daigne11}. In this case, the real $\s_{15}$ should be $\ll 1$, so that $(1+\s_{15}) \simeq 1$. If the derived $(1+\s_{15})$ is larger than 1, it suggests that significant non-thermal emission is generated through ICMART \citep{zhangyan11}.

We again consider the six regimes for the non-dissipative photosphere studied in section \ref{sec:no-mag-bottom}. Similar to \cite{pe'er07}, Regimes I and IV ($r_{\rm ph} < r_{\rm ra}$) introduce an additional degeneracy, so that central engine parameters cannot be inferred. We therefore focus on the other four regimes. The criteria for the four regimes based on observations are summarized in Table 2.

For regime II, we have

\begin{eqnarray}
&&1+\s_0= 25.5(1+z)^{4/3}\left(\frac{k T_{\rm ob}}{50~{\rm keV}}\right)^{4/3}\left(\frac{F_{\rm BB}}{10^{-8}~{\rm erg~s^{-1}cm^{-2}}}\right)^{-1/3}r_{0,9}^{2/3}f_{\rm th,-1}^{-1}f_{\gamma}^{-1}d_{L,28}^{-2/3},\nonumber\\
&&\eta=74.8(1+z)^{11/12}\left(\frac{k T_{\rm ob}}{50~{\rm keV}}\right)^{11/12}\left(\frac{F_{\rm BB}}{10^{-8}~{\rm erg~s^{-1}cm^{-2}}}\right)^{1/48}r_{0,9}^{5/24}d_{L,28}^{1/24},\nonumber\\
&&r_{\rm ph}=1.78\times 10^{10}~ {\rm cm}(1+z)^{-25/12}\left(\frac{k T_{\rm ob}}{50~{\rm keV}}\right)^{-25/12}\left(\frac{F_{\rm BB}}{10^{-8}~{\rm erg~s^{-1}cm^{-2}}}\right)^{37/48}r_{0,9}^{-7/24}d_{L,28}^{37/24} ,\nonumber\\
&&\g_{\rm ph}=46.4(1+z)^{-1/12}\left(\frac{k T_{\rm ob}}{50~{\rm keV}}\right)^{-1/12}\left(\frac{F_{\rm BB}}{10^{-8}~{\rm erg~s^{-1}cm^{-2}}}\right)^{13/48}r_{0,9}^{-7/24}d_{L,28}^{13/24}.\nonumber\\
&&1+\s_{\rm ph}=41.2(1+z)^{7/3}\left(\frac{k T_{\rm ob}}{50~{\rm keV}}\right)^{7/3}\left(\frac{F_{\rm BB}}{10^{-8}~{\rm erg~s^{-1}cm^{-2}}}\right)^{-7/12}r_{0,9}^{7/6}f_{\rm th,-1}^{-1}f_{\gamma}^{-1}d_{L,28}^{-7/6}.\nonumber\\
&&1+\s_{r_{15}}=1.08(1+z)^{59/36}\left(\frac{k T_{\rm ob}}{50~{\rm keV}}\right)^{59/36}\left(\frac{F_{\rm BB}}{10^{-8}~{\rm erg~s^{-1}cm^{-2}}}\right)^{-47/144}r_{0,9}^{77/72}f_{\rm th,-1}^{-1}f_{\gamma}^{-1}d_{L,28}^{-47/72}.\nonumber\\
\end{eqnarray}

For regime III and VI, we have 

\begin{eqnarray}
&&1+\s_0= 5.99(1+z)^{4/3}\left(\frac{k T_{\rm ob}}{30~{\rm keV}}\right)^{4/3}\left(\frac{F_{\rm BB}}{10^{-7}~{\rm erg~s^{-1}cm^{-2}}}\right)^{-1/3}r_{0,9}^{2/3}f_{\rm th,-1}^{-1}f_{\gamma}^{-1}d_{L,28}^{-2/3},\nonumber\\
&&\eta= 20.3(1+z)^{-5/6}\left(\frac{k T_{\rm ob}}{30~{\rm keV}}\right)^{-5/6}\left(\frac{F_{\rm BB}}{10^{-7}~{\rm erg~s^{-1}cm^{-2}}}\right)^{11/24}r_{0,9}^{-2/3}f_{\rm th,-1}^{3/4}f_{\gamma}^{3/4}d_{L,28}^{11/12},\nonumber\\
&&r_{\rm ph}=4.09\times 10^{11}~ {\rm cm}(1+z)^{-3/2}\left(\frac{k T_{\rm ob}}{30~{\rm keV}}\right)^{-3/2}\left(\frac{F_{\rm BB}}{10^{-7}~{\rm erg~s^{-1}cm^{-2}}}\right)^{5/8}f_{\rm th,-1}^{-1/4}f_{\gamma}^{-1/4}d_{L,28}^{5/4} ,\nonumber\\
&&\g_{\rm ph}=121.3(1+z)^{1/2}\left(\frac{k T_{\rm ob}}{30~{\rm keV}}\right)^{1/2}\left(\frac{F_{\rm BB}}{10^{-7}~{\rm erg~s^{-1}cm^{-2}}}\right)^{1/8}f_{\rm th,-1}^{-1/4}f_{\gamma}^{-1/4}d_{L,28}^{1/4}.\nonumber\\
\end{eqnarray}

For regime V, we have

\begin{eqnarray}
&&1+\s_0=6.43(1+z)^{4/3}\left(\frac{k T_{\rm ob}}{10~{\rm keV}}\right)^{4/3}\left(\frac{F_{\rm BB}}{10^{-9}~{\rm erg~s^{-1}cm^{-2}}}\right)^{-1/3}r_{0,9}^{2/3}f_{\rm th,-1}^{-1}f_{\gamma}^{-1}d_{L,28}^{-2/3},\nonumber\\
&&\eta= 105.0(1+z)^{7/6}\left(\frac{k T_{\rm ob}}{10~{\rm keV}}\right)^{7/6}\left(\frac{F_{\rm BB}}{10^{-9}~{\rm erg~s^{-1}cm^{-2}}}\right)^{5/24}r_{0,9}^{1/12}f_{\rm th,-1}^{1/2}f_{\gamma}^{1/2}d_{L,28}^{5/12},\nonumber\\
&&r_{\rm ph}=4.62\times 10^{10}~ {\rm cm}(1+z)^{-13/6}\left(\frac{k T_{\rm ob}}{10~{\rm keV}}\right)^{-13/6}\left(\frac{F_{\rm BB}}{10^{-9}~{\rm erg~s^{-1}cm^{-2}}}\right)^{17/24}r_{0,9}^{-1/4}f_{\rm th,-1}^{-1/6}f_{\gamma}^{-1/6}d_{L,28}^{17/12} ,\nonumber\\
&&\g_{\rm ph}=15.3(1+z)^{-1/6}\left(\frac{k T_{\rm ob}}{10~{\rm keV}}\right)^{-1/6}\left(\frac{F_{\rm BB}}{10^{-9}~{\rm erg~s^{-1}cm^{-2}}}\right)^{5/24}r_{0,9}^{-1/4}f_{\rm th,-1}^{-1/6}f_{\gamma}^{-1/6}d_{L,28}^{5/12}.\nonumber\\
&&1+\s_{\rm ph}=44.2(1+z)^{8/3}\left(\frac{k T_{\rm ob}}{10~{\rm keV}}\right)^{8/3}\left(\frac{F_{\rm BB}}{10^{-9}~{\rm erg~s^{-1}cm^{-2}}}\right)^{-1/3}r_{0,9}f_{\rm th,-1}^{-1/3}f_{\gamma}^{-1/3}d_{L,28}^{-2/3}.\nonumber\\
&&1+\s_{r_{15}}=1.59(1+z)^{35/18}\left(\frac{k T_{\rm ob}}{10~{\rm keV}}\right)^{35/18}\left(\frac{F_{\rm BB}}{10^{-9}~{\rm erg~s^{-1}cm^{-2}}}\right)^{-7/72}r_{0,9}^{11/12}f_{\rm th,-1}^{-7/18}f_{\gamma}^{-7/18}d_{L,28}^{-7/36}.\nonumber\\
\end{eqnarray}

Notice that regime VI has the identical scalings as regime III.

\section{Case study: GRB 110721A}

%With results in section 4, one can diagnose GRB central engine properties with observed prompt spectrum information.  We take GRB 110721A as an example. 
GRB 110721A was jointly detected by the {\em Fermi} GBM and LAT. \cite{axelsson12} reported the time-dependent spectral evolution of GRB 110721A and suggested that the time-resolved emission spectra are best modeled with a combination of a Band function and a blackbody component. Based on a candidate optical counterpart \citep{greiner11}, \cite{berger11} suggested two possible redshifts, $z = 0.382$ or $z = 3.512$, with the former preferred. 

\begin{figure}[h!!!]
\subfigure[]{
    \label{fig:subfig:a} %% label for first subfigure
    \includegraphics[width=2.0in]{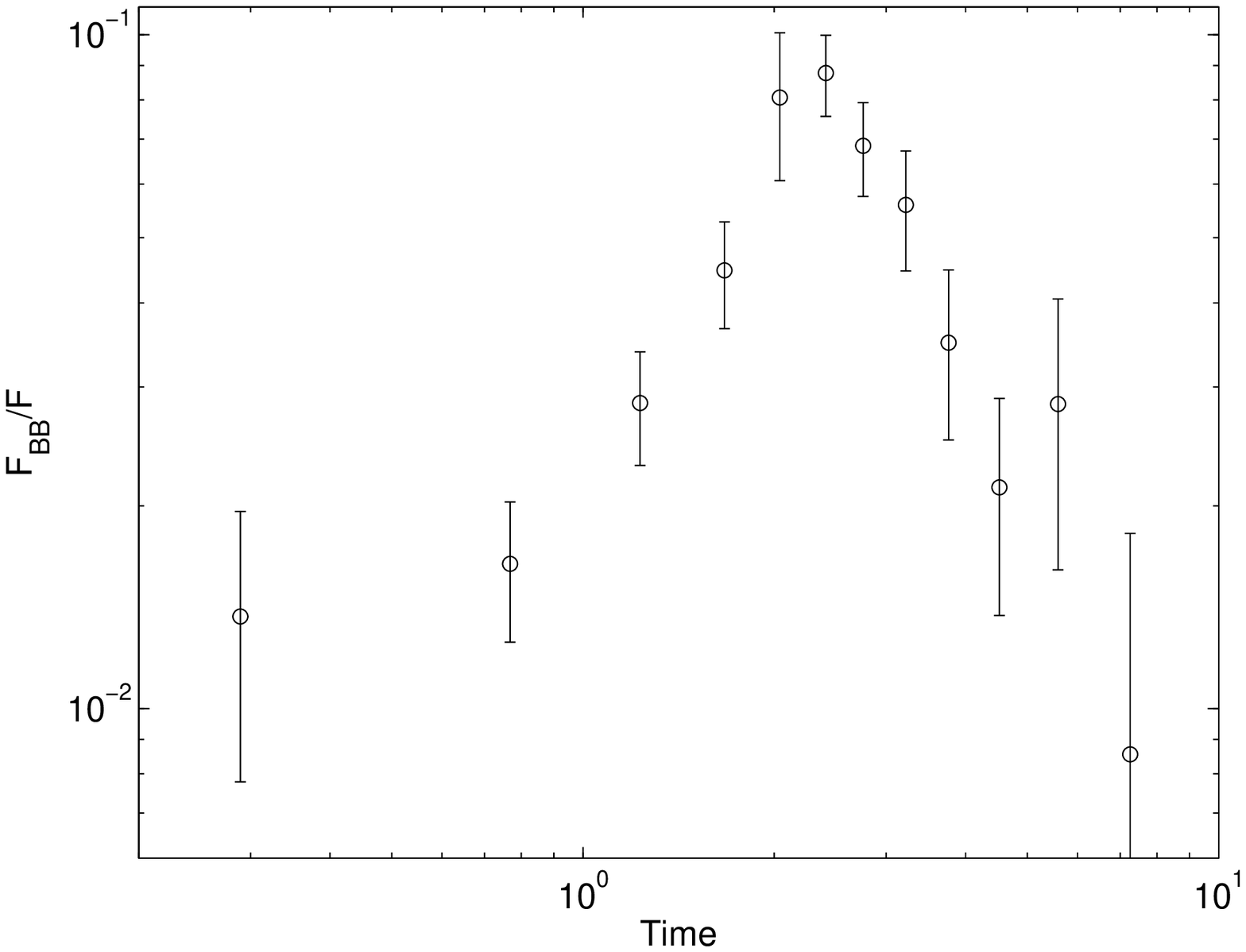}}
    \subfigure[]{
\label{fig:subfig:b} %% label for first subfigure
    \includegraphics[width=2.0in]{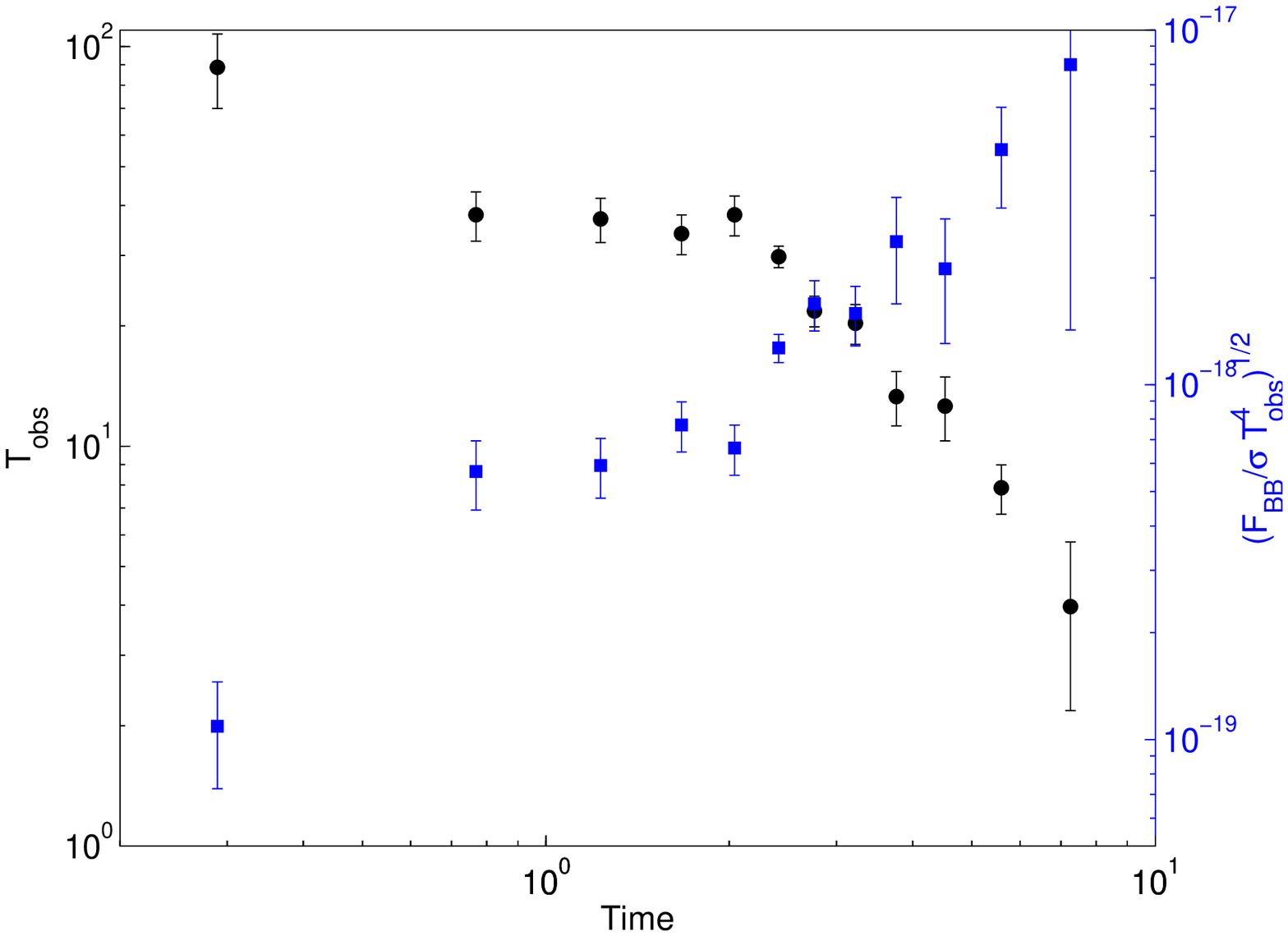}}
    \subfigure[]{
    \label{fig:subfig:c} %% label for first subfigure
    \includegraphics[width=2.0in]{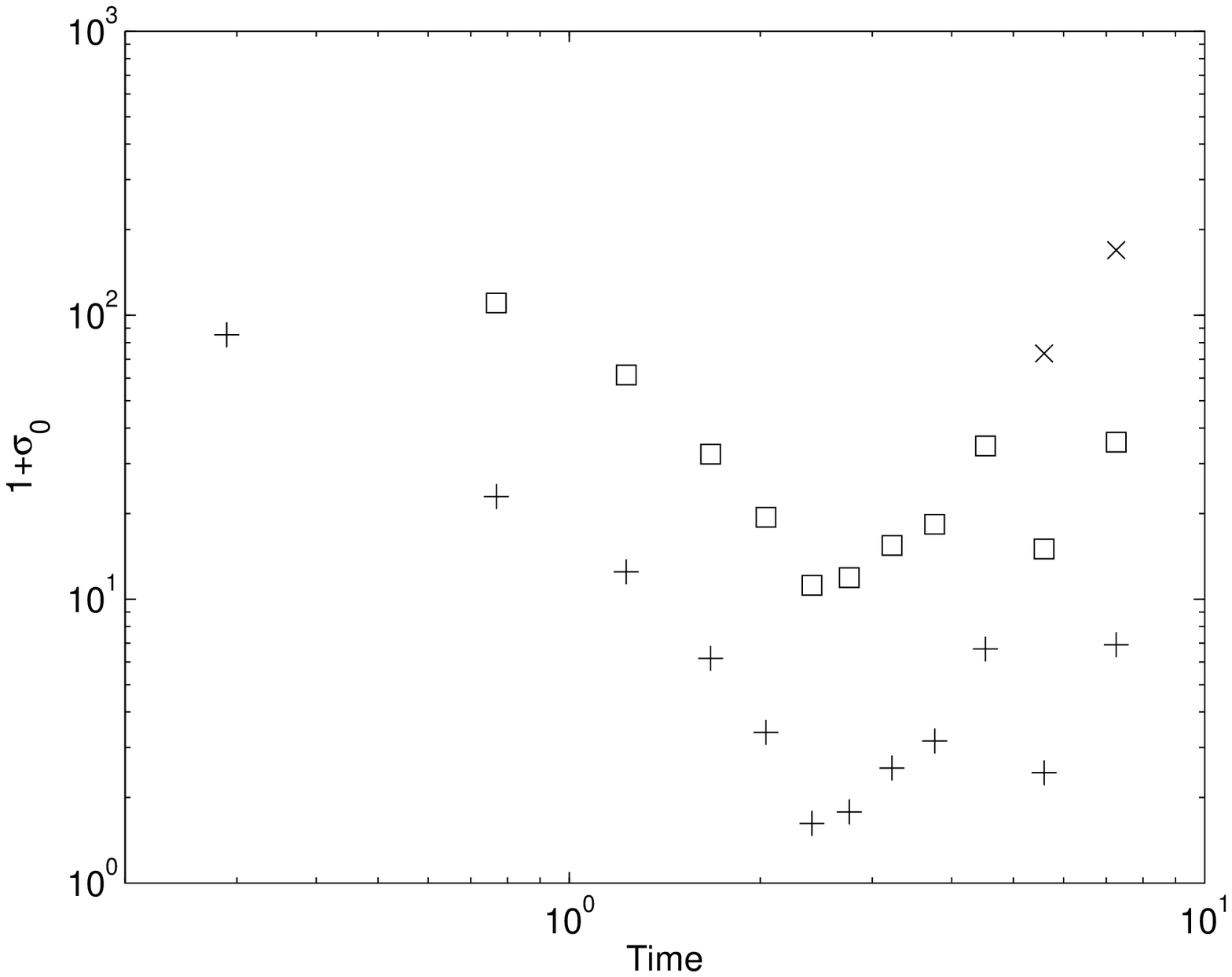}}
    \subfigure[]{
    \label{fig:subfig:d} %% label for first subfigure
    \includegraphics[width=2.0in]{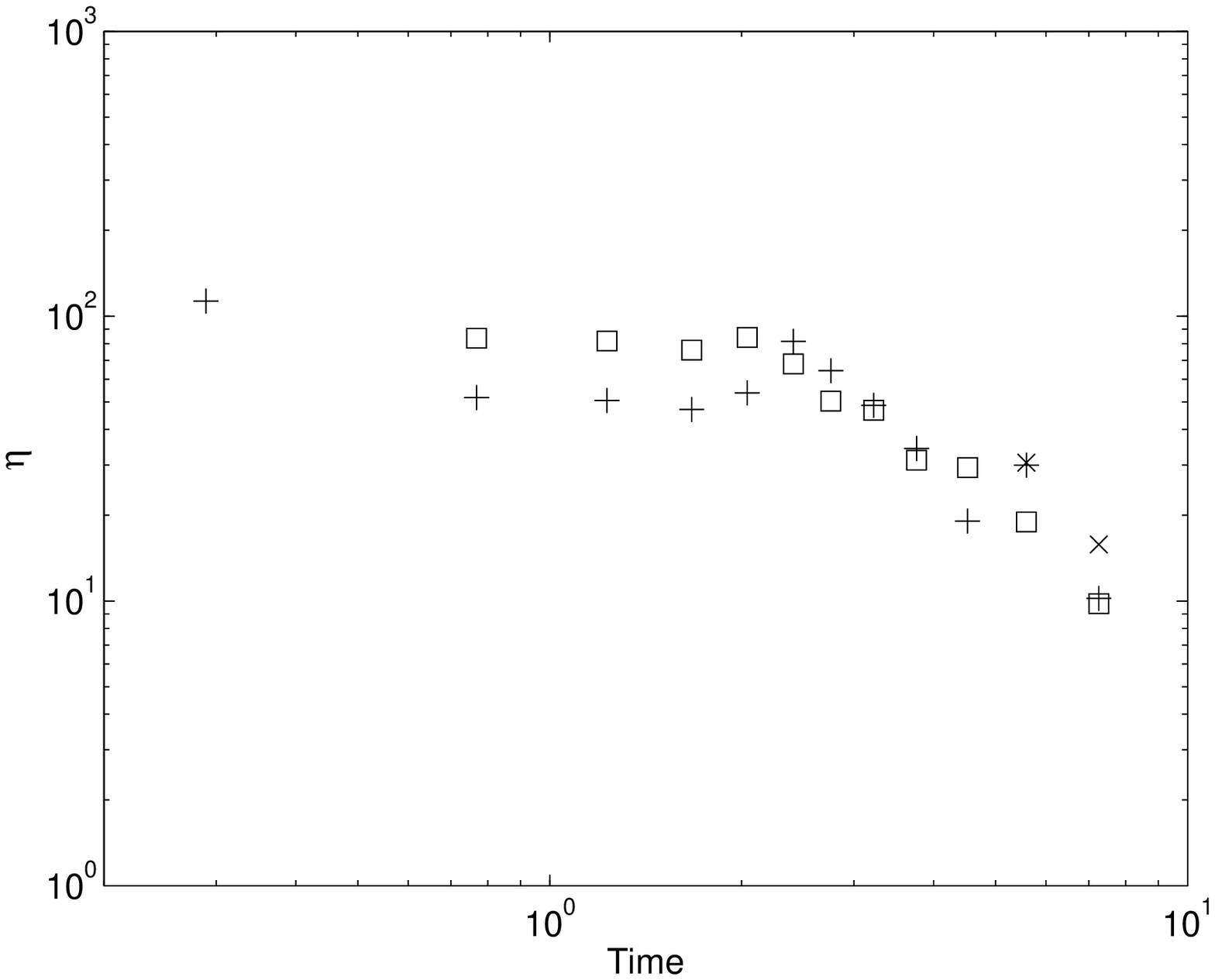}}
    \subfigure[]{
    \label{fig:subfig:e} %% label for first subfigure
    \includegraphics[width=2.0in]{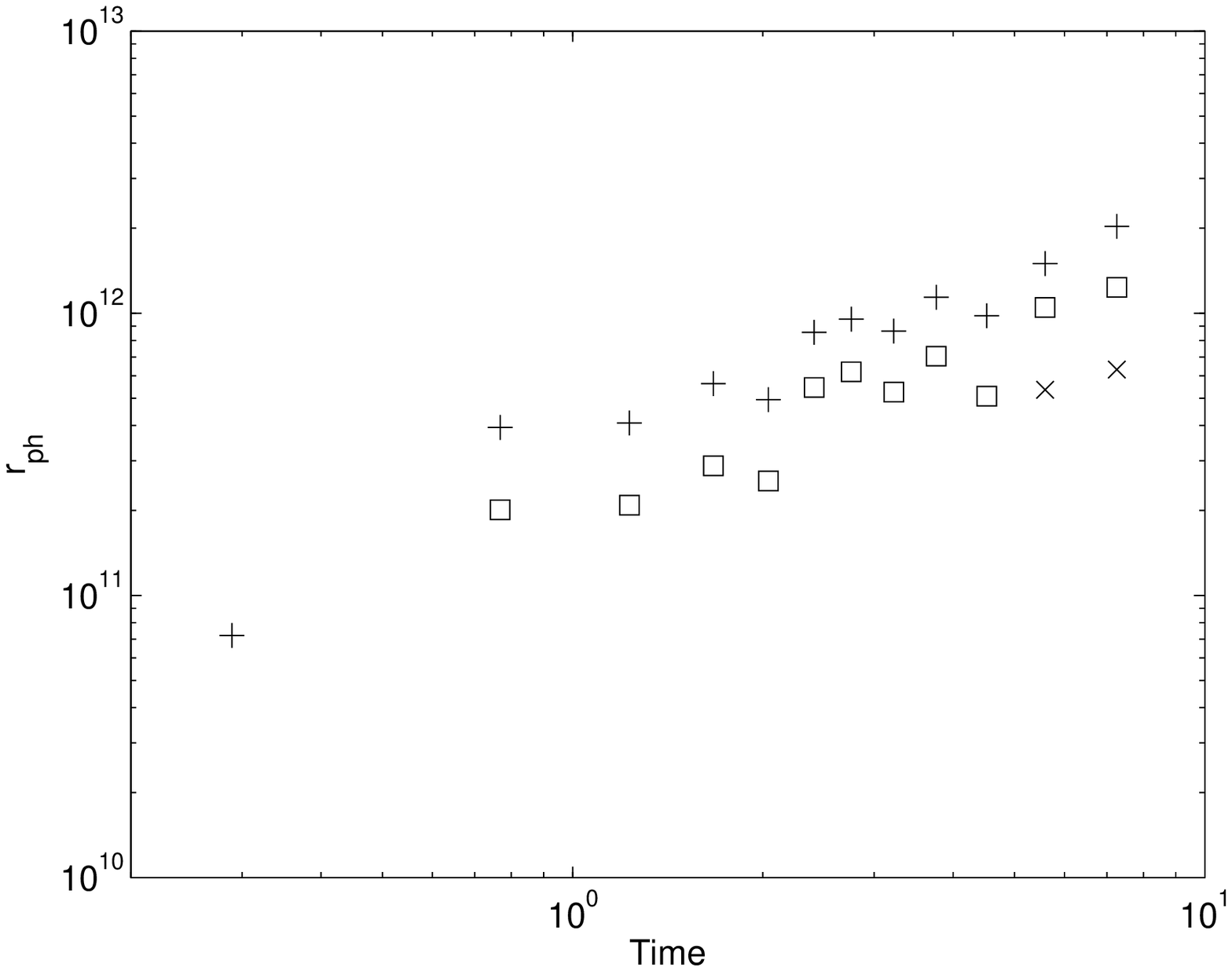}}
    \subfigure[]{
    \label{fig:subfig:f} %% label for first subfigure
    \includegraphics[width=2.0in]{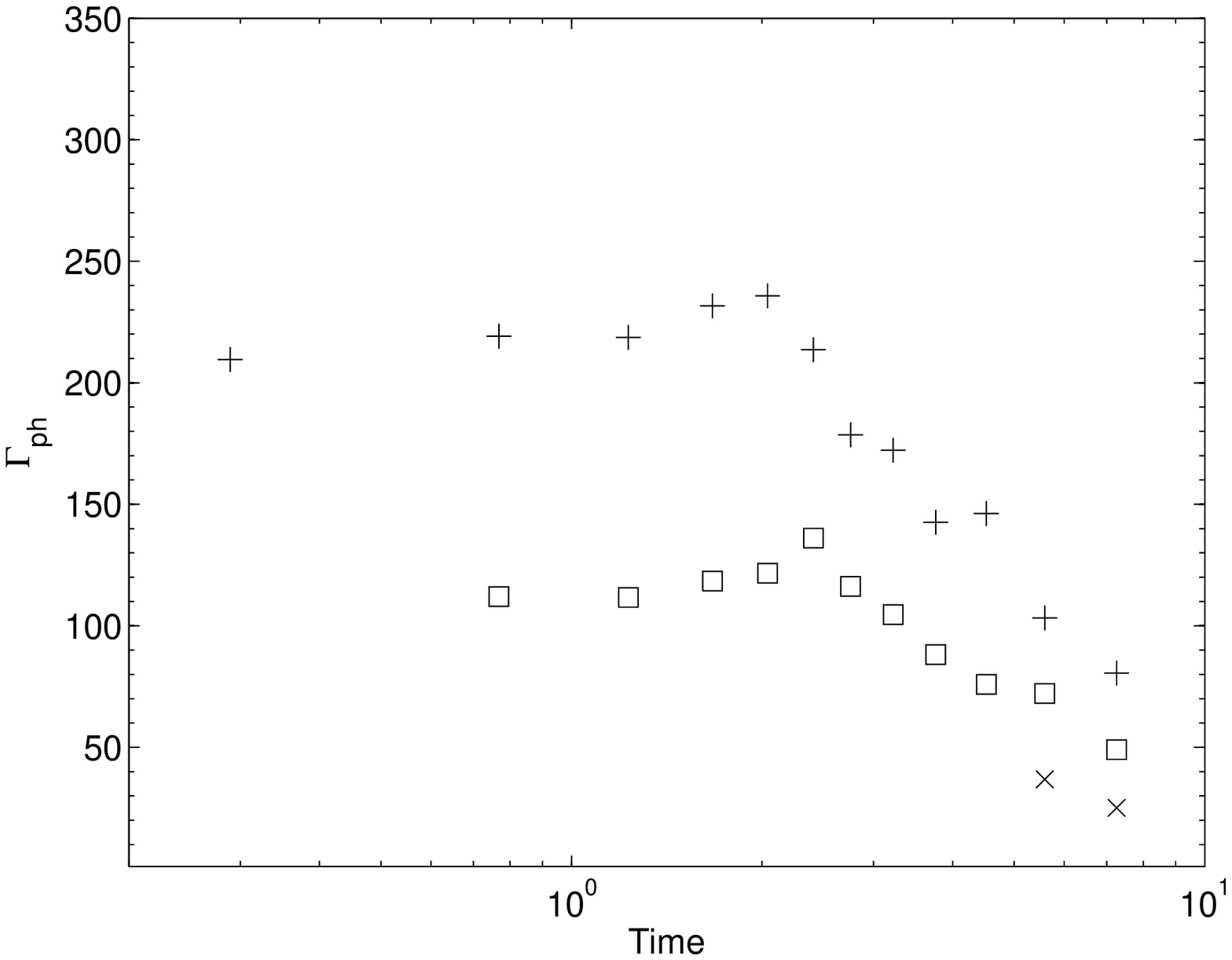} }
    \subfigure[]{
    \label{fig:subfig:g} %% label for first subfigure
    \includegraphics[width=2.0in]{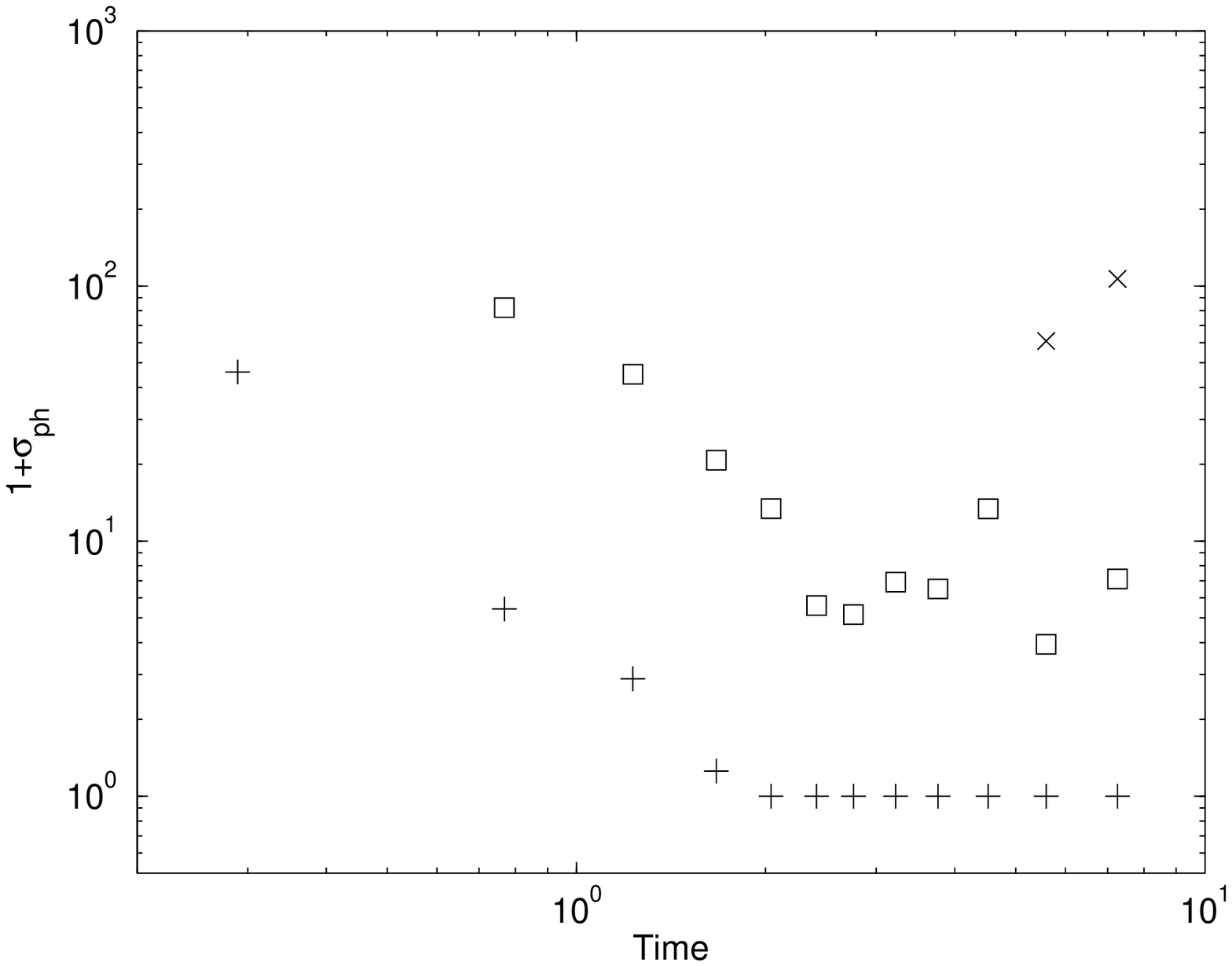} }
    \subfigure[]{
    \label{fig:subfig:h} %% label for first subfigure
    \includegraphics[width=2.0in]{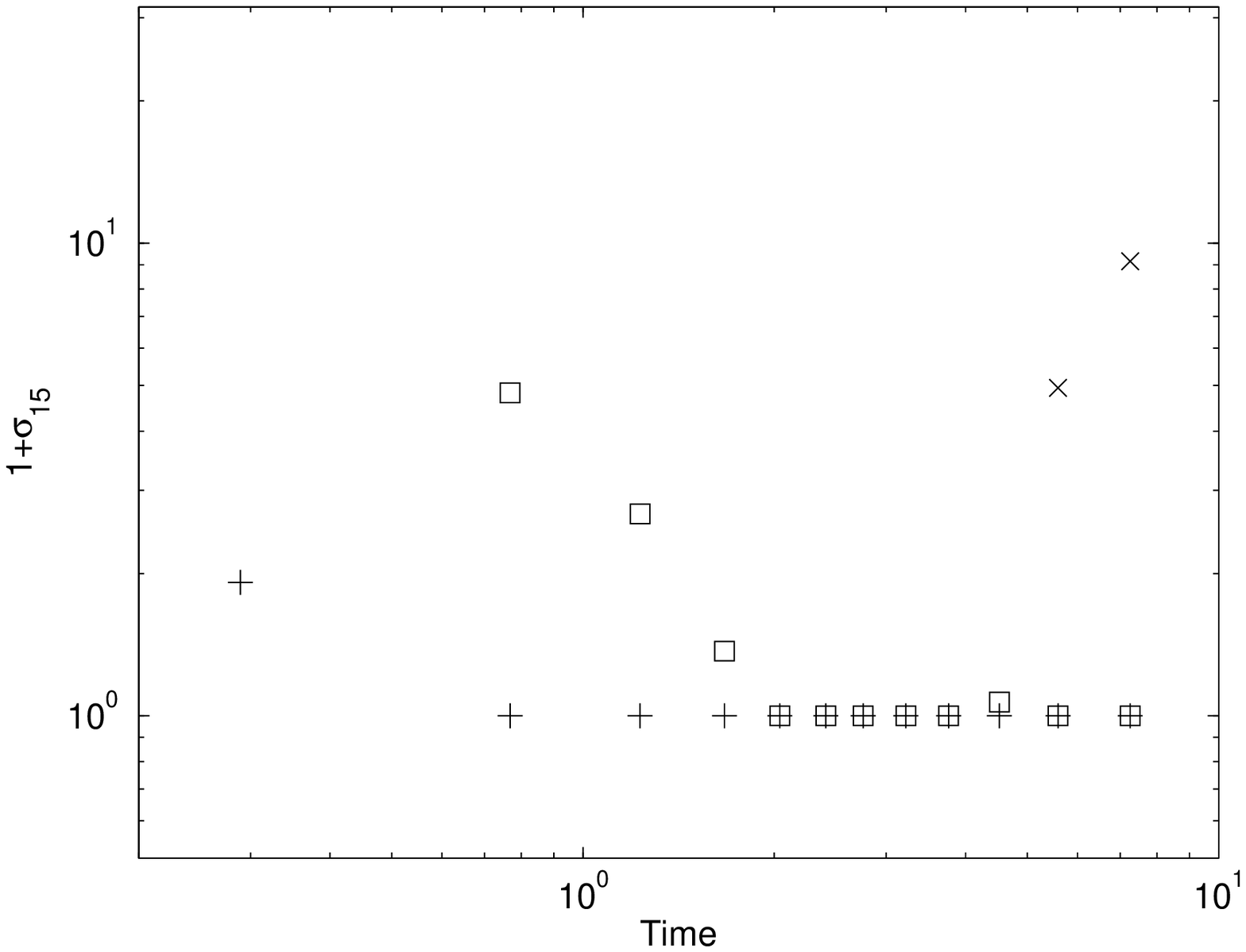} }
      \caption{A case study for GRB 110721A. The observed $F_{\rm BB}/F_{\rm ob}$ (a) and the observed $T_{\rm BB}$ and $(F_{\rm BB}/\sigma T_{\rm ob}^4)^{1/2}$ (b) evolution \citep{iyyani13}, along with the derived parameters of a non-dissipative photosphere and their evolution: $1+\s_0$ (c), $\eta$ (d), $r_{\rm ph}$ (e), $\g_{\rm ph}$ (f), $(1+\s_{\rm ph})$ (g), and $(1+\s_{15})$ (h). The plus, square and cross symbols denote the cases with $r_0=10^8~{\rm cm},~10^{9}~{\rm cm}$ and $10^{10}~{\rm cm}$, respectively.}
           \label{fig:110721a}
            \end{figure}

\cite{iyyani13} analyzed the time-resolved data of GRB 110721A and presented the time-dependent properties (including $f_{\rm th}=F_{\rm BB}/F_{\rm ob}$, panel (a) of Fig.\ref{fig:110721a}; $T_{\rm ob}$, and $(F_{\rm BB}/\sigma T_{ob}^4)^{1/2}$, panel (b) of Fig.\ref{fig:110721a}). Based on the diagnostic method of \cite{pe'er07}, they derived $\eta(t)$ and $r_0(t)$. Some uncomfortable conclusions were obtained: First, $\eta(t)$ decreases with time. This is at odds with the IS model, which demands a time-increasing $\eta(t)$ to make strong ISs and efficient non-thermal emission. Second, $r_0$ was found to increase by more than two orders of magnitude early on and later decrease by near one order of magnitude. It is hard to imagine any realistic central engine that may change its size with such a large amplitude in such a short period of time.

By applying our top-down diagnostic method, the data can be naturally explained (Fig.\ref{fig:110721a}). We find that in all time bins, one has $(1+\s_0) \gg 1$, so that the \cite{pe'er07} approach cannot be applied. The variation of the thermal emission properties in the time-resolved spectra is a result of varying $(\eta, \s_0)$ pair as a function of time at the central engine. This is a more reasonable conclusion as compared with the varying $r_0$ result. The GRB central engine is highly erratic, so that it is possible that the dimensionless entropy and magnetization can vary noticeably with time. Our inferred parameters depend on the assumed constant $r_0$. In Fig.\ref{fig:110721a} (panels c,d), we present the results of $(1+\s_0)$ and $\eta$ as a function of time for three assumed $r_0$: $10^8$ (plus), $10^9$ (square), and $10^{10}$ (cross) cm. 
We also plot the photosphere radius $r_{\rm ph}$ and photosphere Lorentz factor $\Gamma_{\rm ph}$ as a function of $t$ for all the cases (panels (e,f) of Fig.\ref{fig:110721a}). It is interesting to see that the derived $(1+\s_0)$ initially increases with time, which is consistent with the expectation of some central engine models (e.g. \citealt{metzger11}, W.-H. Lei et al. 2014, in preparation). It is also interesting to note that $\Gamma_{\rm ph}$ initially rises with time, in contrast to the case of the pure fireball model \citep{iyyani13}. This is a more natural picture for both IS and ICMART scenarios. For this case, $(1+\s_{15})$ is found to be above unity for some time bins. This suggests that ICMART rather than IS is the mechanism to power the non-thermal emission for at least some, and probably all time bins (even if $\sigma_{15}$ is smaller, but not much smaller than unity, IS is still inefficient, and ICMART can enhance energy dissipation). Based on the results in Section \ref{sec:top-down}, we find that $(1+\s_{15})$ tends to be larger for a larger $r_0$. This is confirmed from the analysis of this burst.

\section{Conclusions and Discussion}
\label{sec:discussion}

\begin{figure}[h!!!]
\subfigure[]{
    \label{fig:subfig:a} %% label for first subfigure
    \includegraphics[width=2.0in]{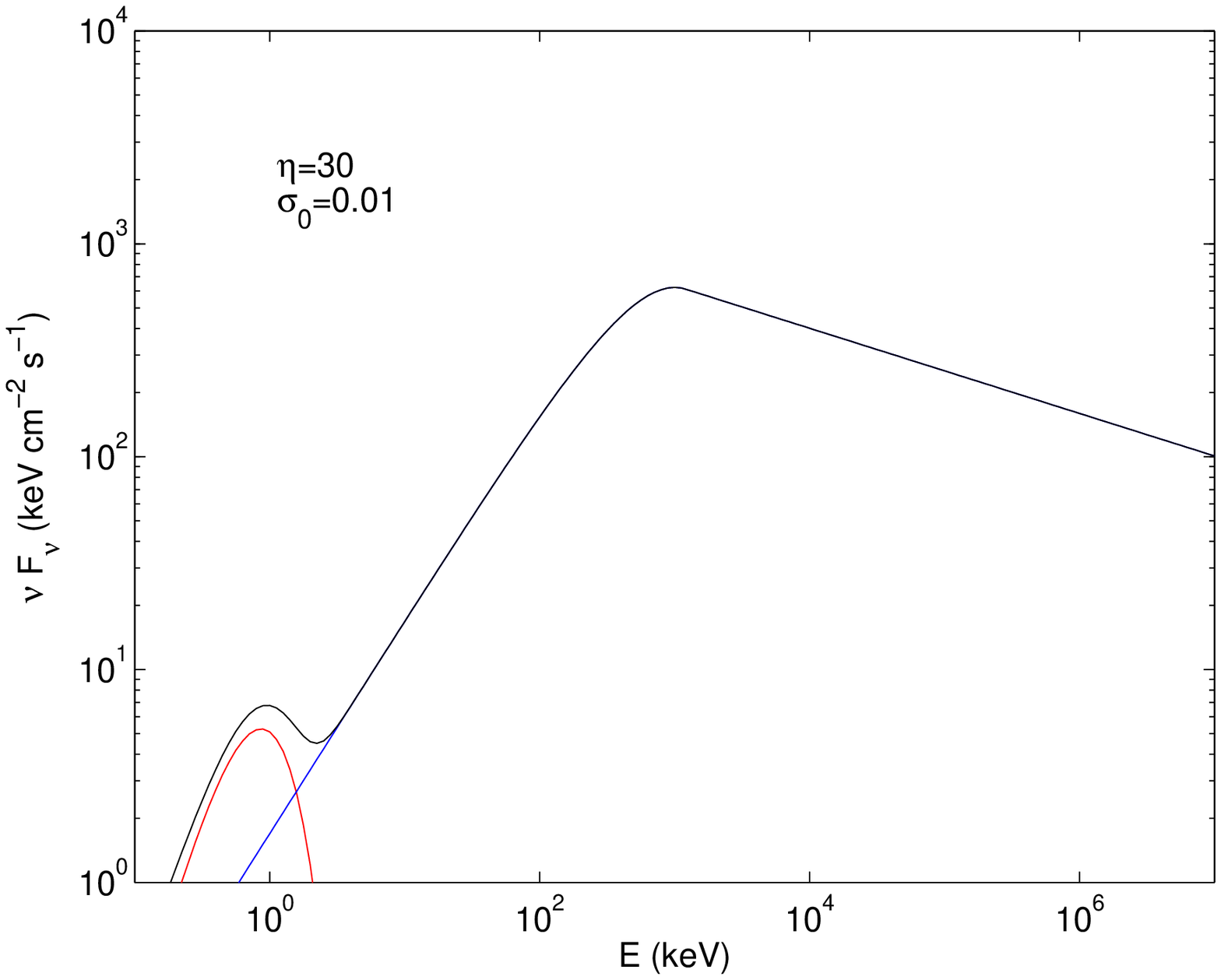}}
    \subfigure[]{
\label{fig:subfig:b} %% label for first subfigure
    \includegraphics[width=2.0in]{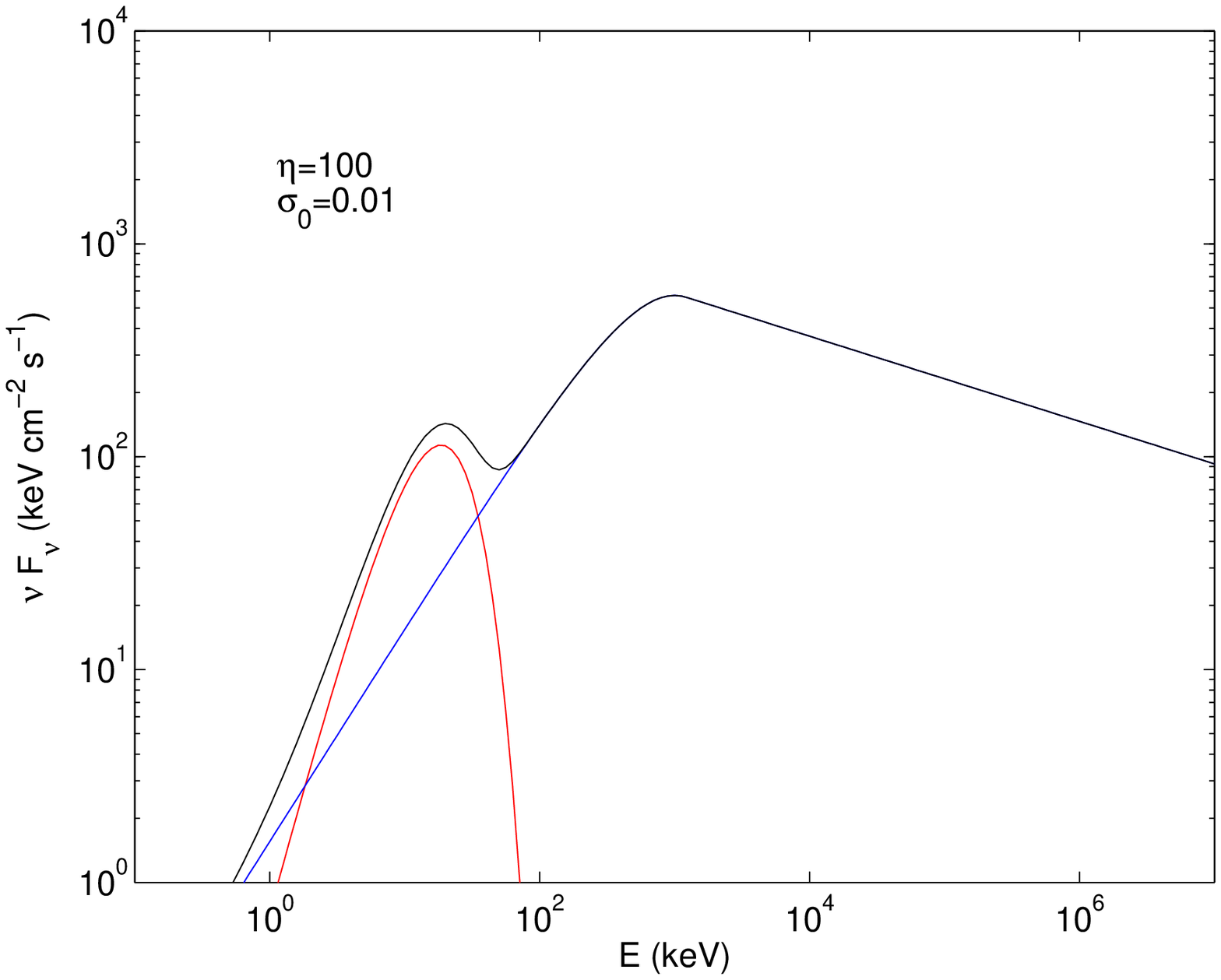}}
    \subfigure[]{
    \label{fig:subfig:c} %% label for first subfigure
    \includegraphics[width=2.0in]{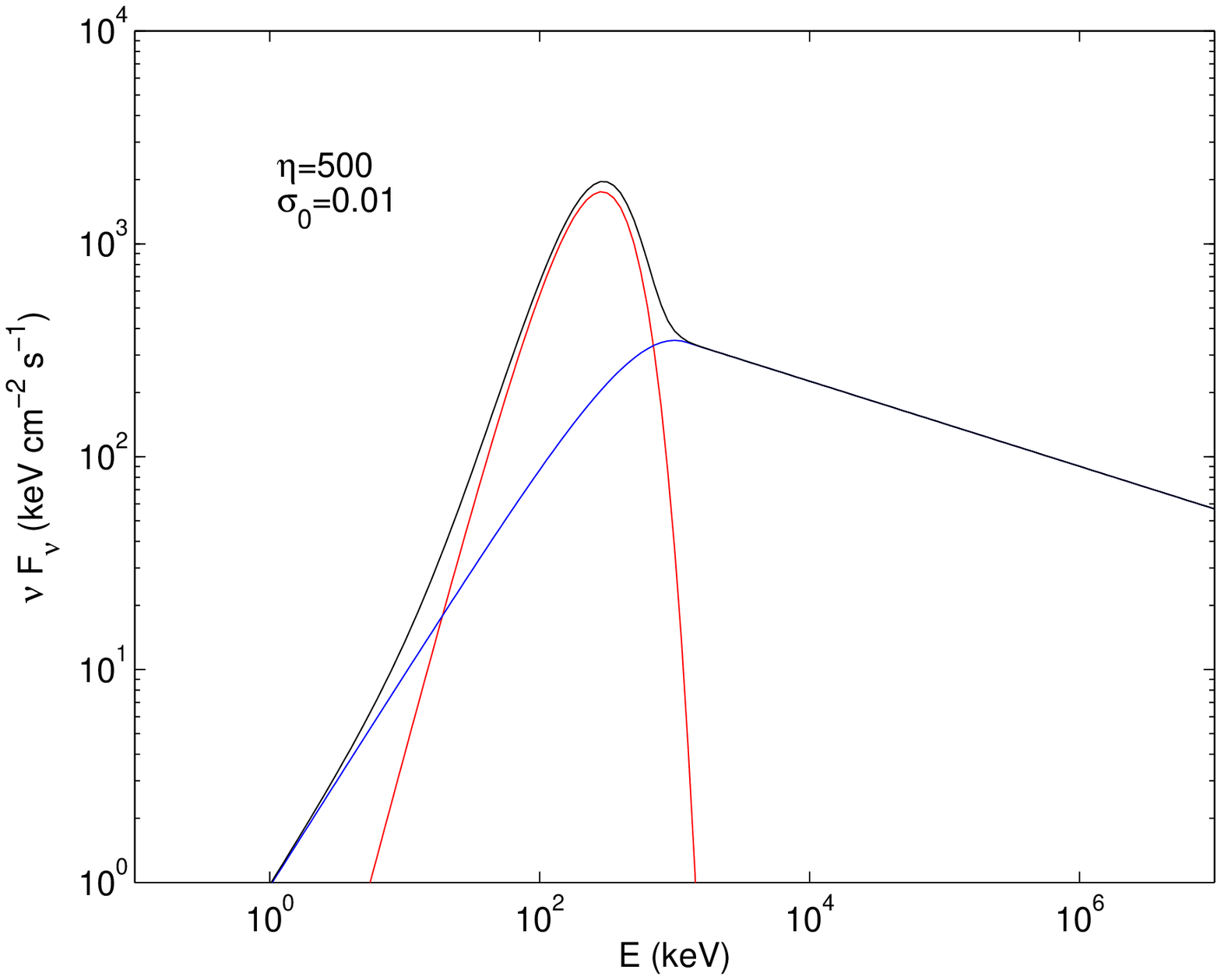}}
    \subfigure[]{
    \label{fig:subfig:d} %% label for first subfigure
    \includegraphics[width=2.0in]{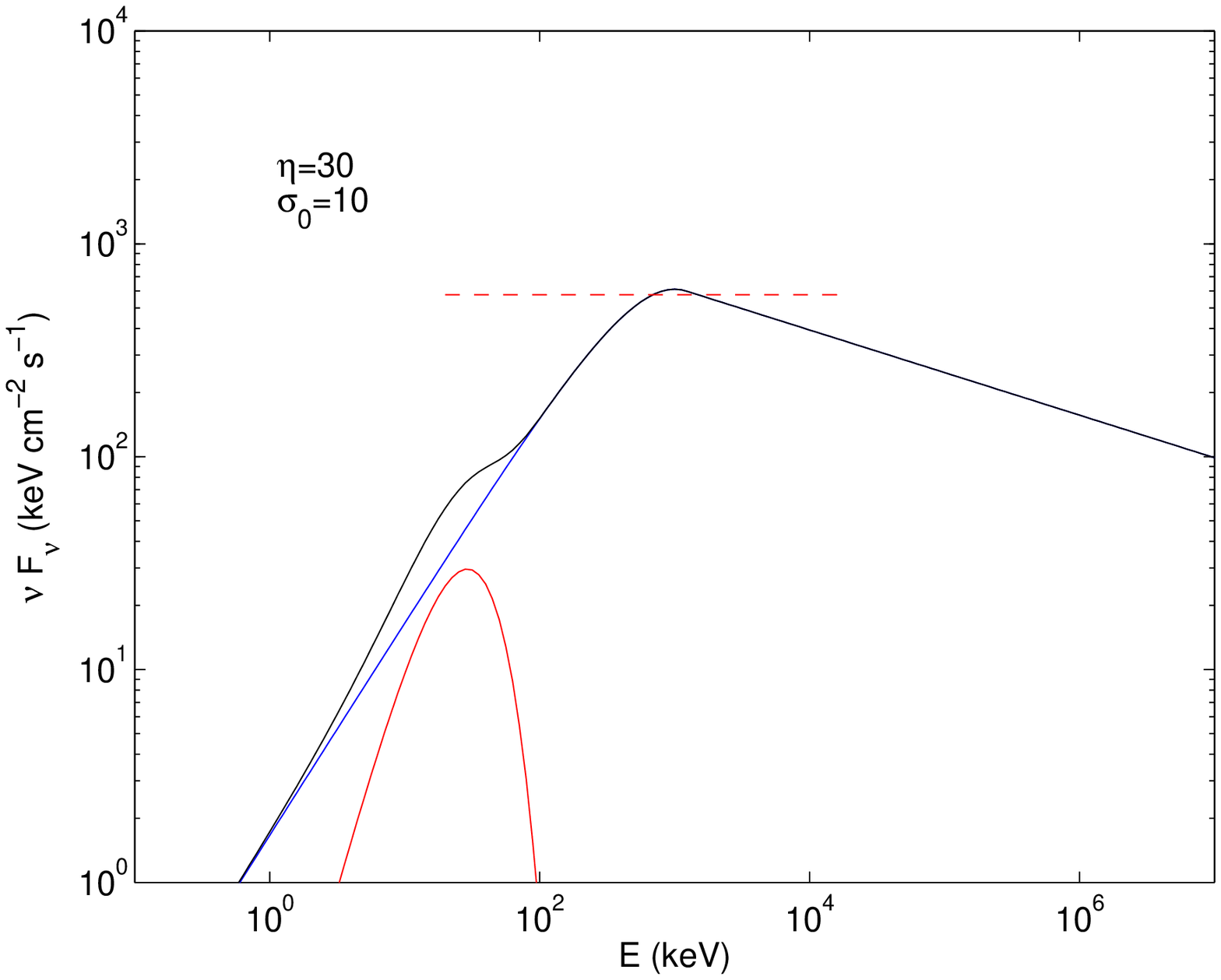}}
    \subfigure[]{
    \label{fig:subfig:a} %% label for first subfigure
    \includegraphics[width=2.0in]{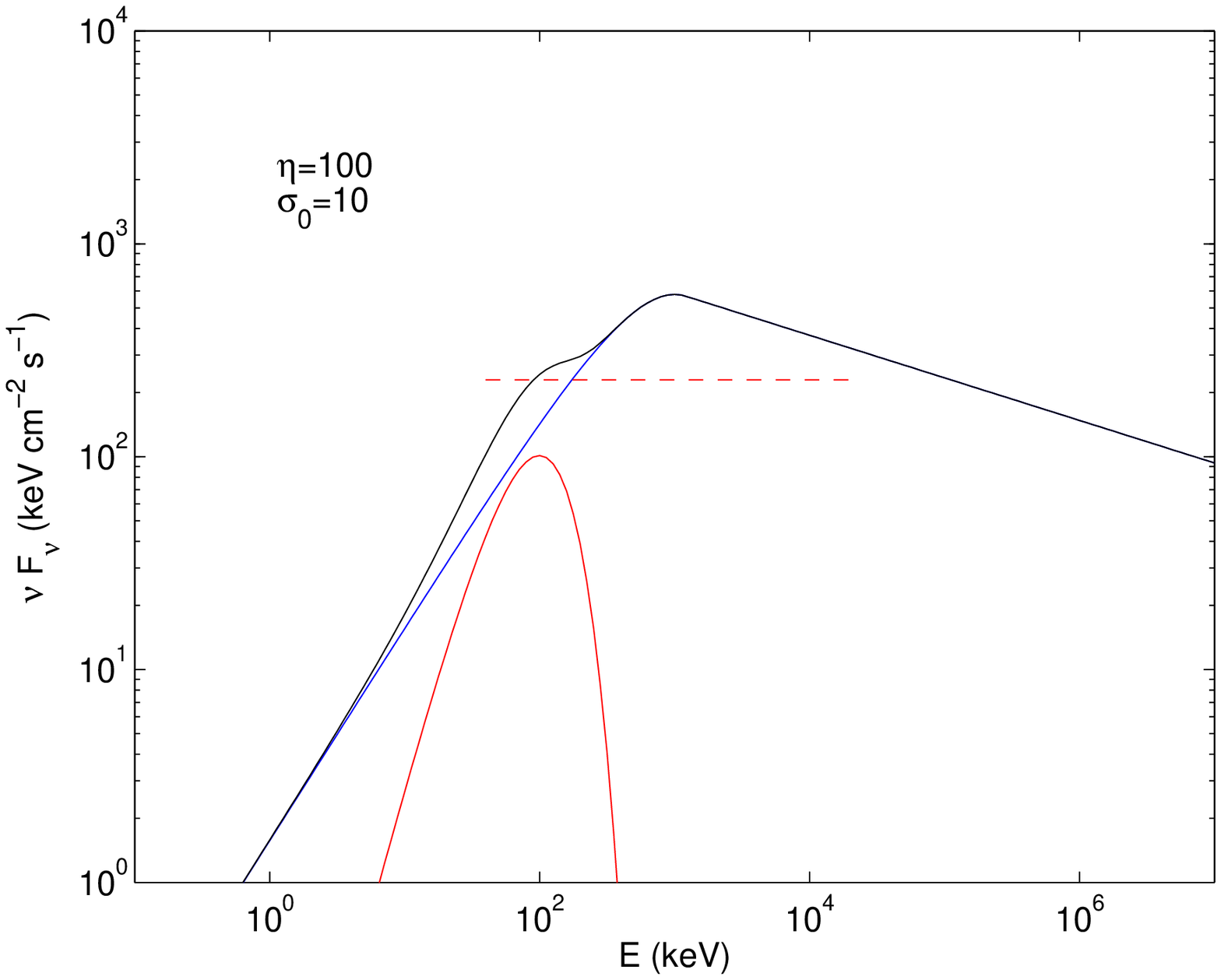}}
    \subfigure[]{
    \label{fig:subfig:a} %% label for first subfigure
    \includegraphics[width=2.0in]{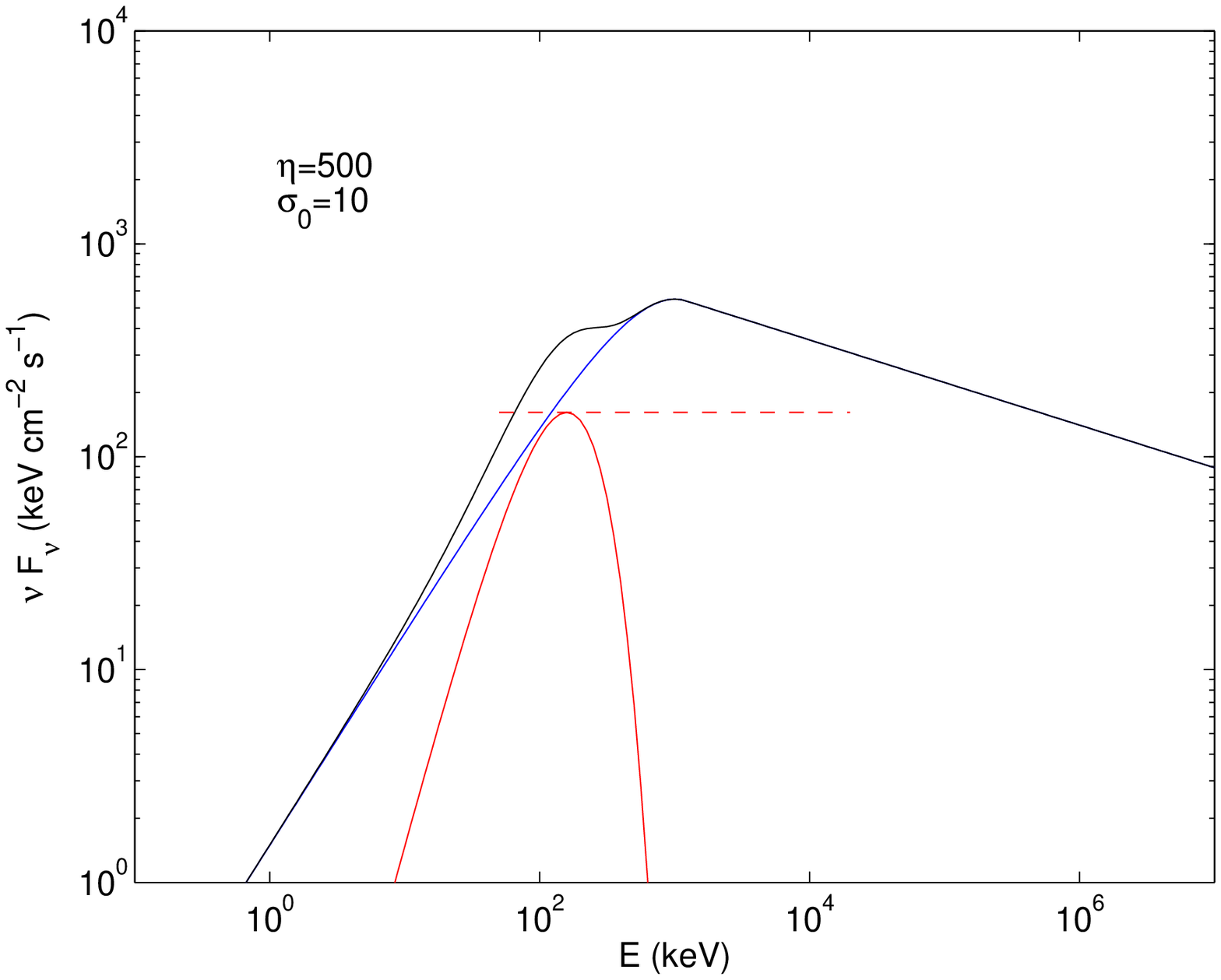}}
    \subfigure[]{
    \label{fig:subfig:a} %% label for first subfigure
    \includegraphics[width=2.0in]{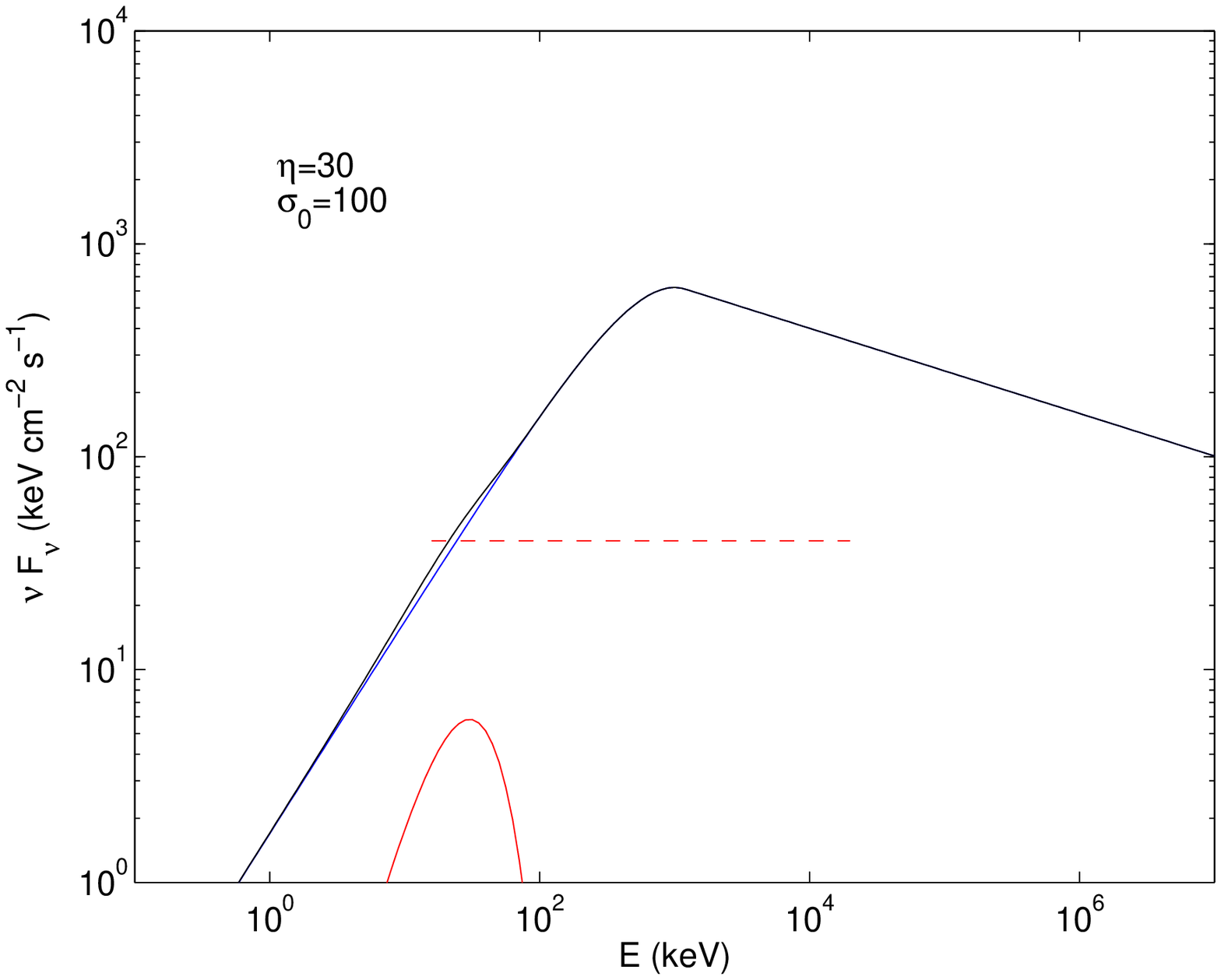}}
    \subfigure[]{
    \label{fig:subfig:a} %% label for first subfigure
    \includegraphics[width=2.0in]{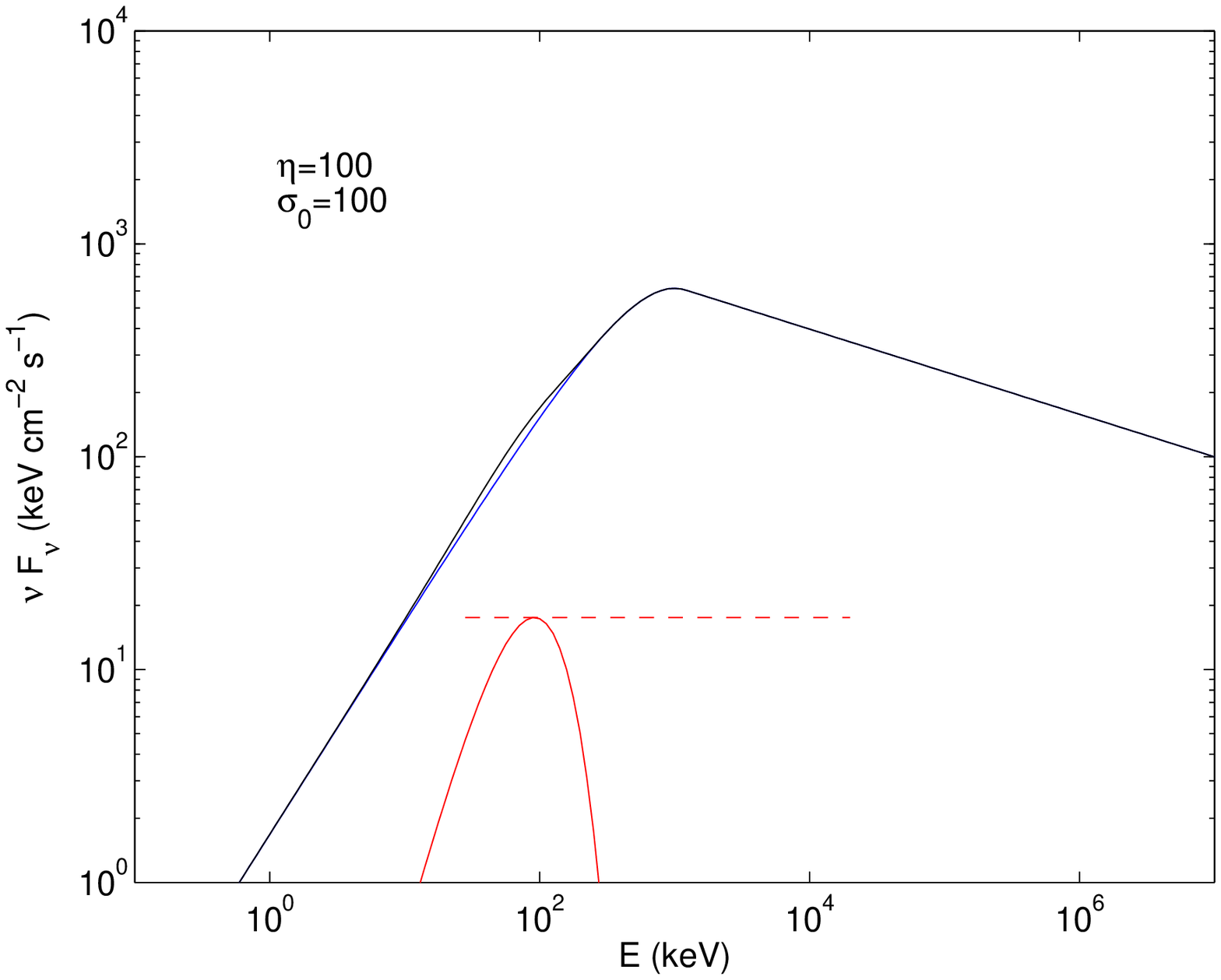}}
    \subfigure[]{
    \label{fig:subfig:a} %% label for first subfigure
    \includegraphics[width=2.0in]{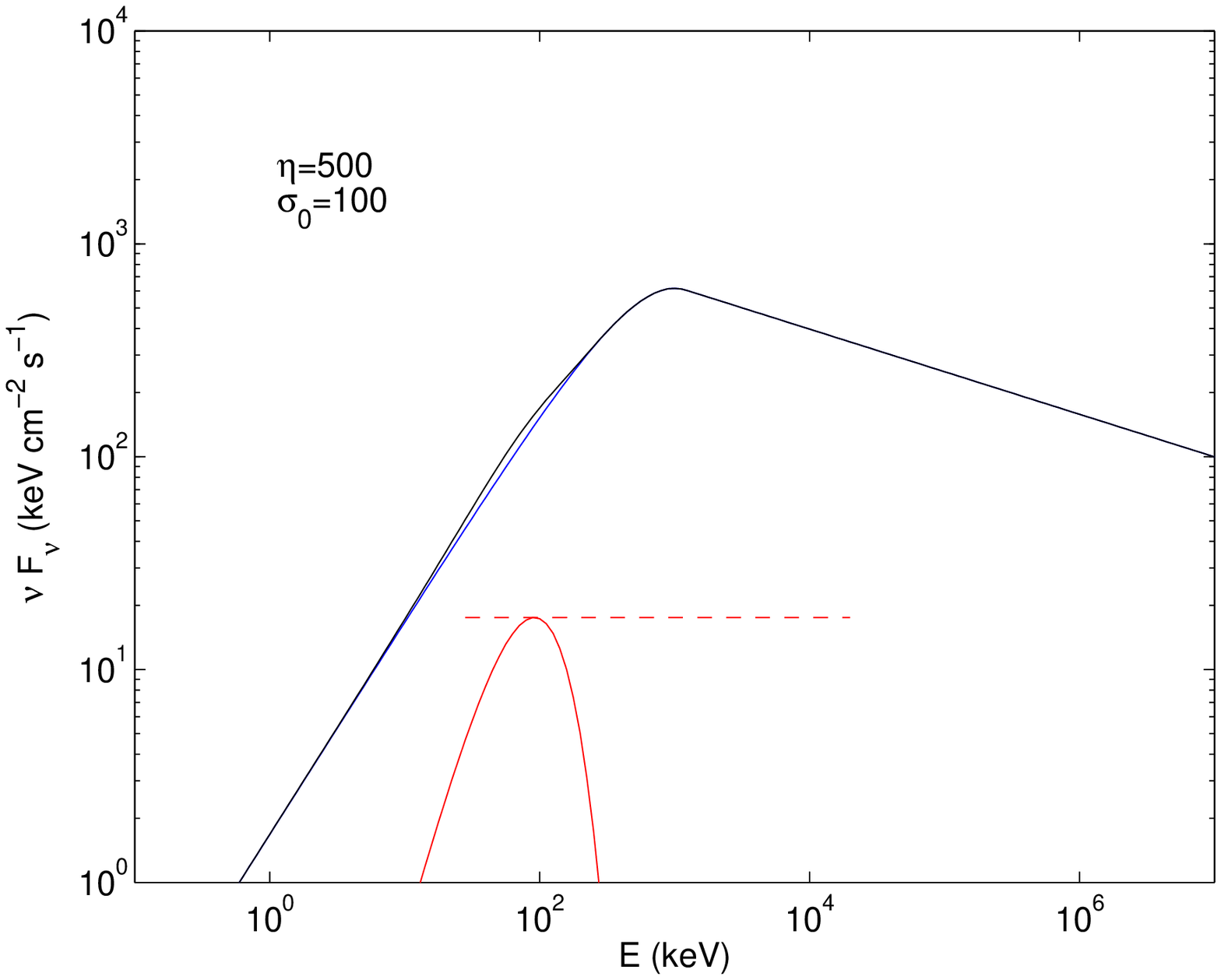}}
      \caption{Example model spectra of GRB prompt emission. Red, blue and black lines are for the thermal (non-dissipative photosphere scenario), non-thermal, and total spectral components, respectively.  Here we adopt $L_w=10^{52} {\rm erg~s^{-1}}$, $r_0=10^9 {\rm cm}$, and $z=1$. Different panels correspond to different combination of $\eta$ and $\s_0$ (as marked in the inset of each panel). The non-thermal radiation efficiency is assumed as 50\%, and a typical Band function shape is adopted. Dashed lines represent the flux and $E_p$ range of photosphere emission for the dissipative photosphere scenario. The lower limit of $E_p$ is calculated assuming the photosphere emission is thermal, and the upper limit is fixed as 20 Mev \citep{beloborodov13}.}
           \label{fig:spectrum}
            \end{figure}

The central engine of GRB jets is most likely a hybrid system with both a hot fireball component and a cold Poynting flux component. In this work, we developed an analytical theory to quantify the properties of the photosphere emission of such a hybrid system. Based on an approximate dynamical evolution model of the hybrid system, we developed a ``bottom-up'' approach to predict the temperature and luminosity of the photosphere emission for arbitrary input parameters, especially $\eta$ and $\s_0$. We consider the cases of both with and without significant magnetic dissipation below the photosphere. We show that a variety of observed GRB prompt emission spectra by {\em Fermi} can be reproduced for the non-dissipative photosphere model, given that $\eta$ and $\s_0$ are allowed to vary in a wide range (Fig.\ref{fig:spectrum}).  We also developed a ``top-down'' approach to diagnose $\eta$ and $\s_0$ within the non-dissipative photosphere model using the observations that show a superposed blackbody component in the GRB spectra. We apply the method to GRB 110721A and draw the conclusion that the central engine of the source as well as its photosphere is highly magnetized, and that the non-thermal emission is likely produced via magnetic dissipation (ICMART) rather than internal shocks. The rapid evolution of the photosphere emission properties is a result of rapid evolution of $(\eta,\s_0)$ pairs rather than rapid evolution of $r_0$ \citep[cf.][]{iyyani13}. We recommend to apply our method to diagnose a large sample of {\em Fermi} GRBs with the detected photosphere component, so as to carry out a statistical analysis of the central engine properties of a large sample of GRBs. This would have interesting implications in diagnosing the composition of GRB jets as well as inferring the mechanisms of GRB jet energy dissipation and radiation.

\begin{figure}[h!!!]
\subfigure[]{
    \label{fig:subfig:a} %% label for first subfigure
    \includegraphics[width=3.0in]{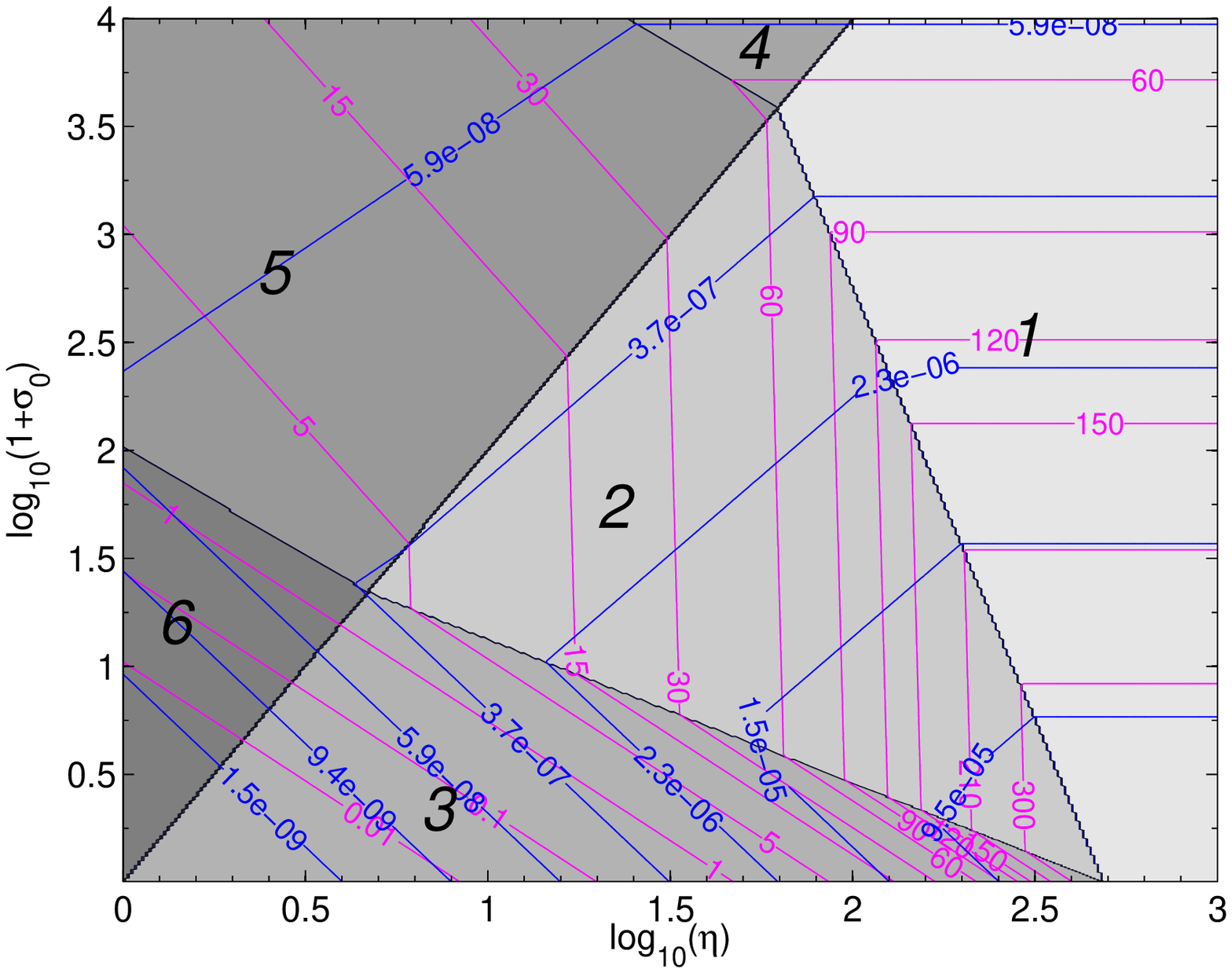}}
    \subfigure[]{
\label{fig:subfig:b} %% label for first subfigure
    \includegraphics[width=3.0in]{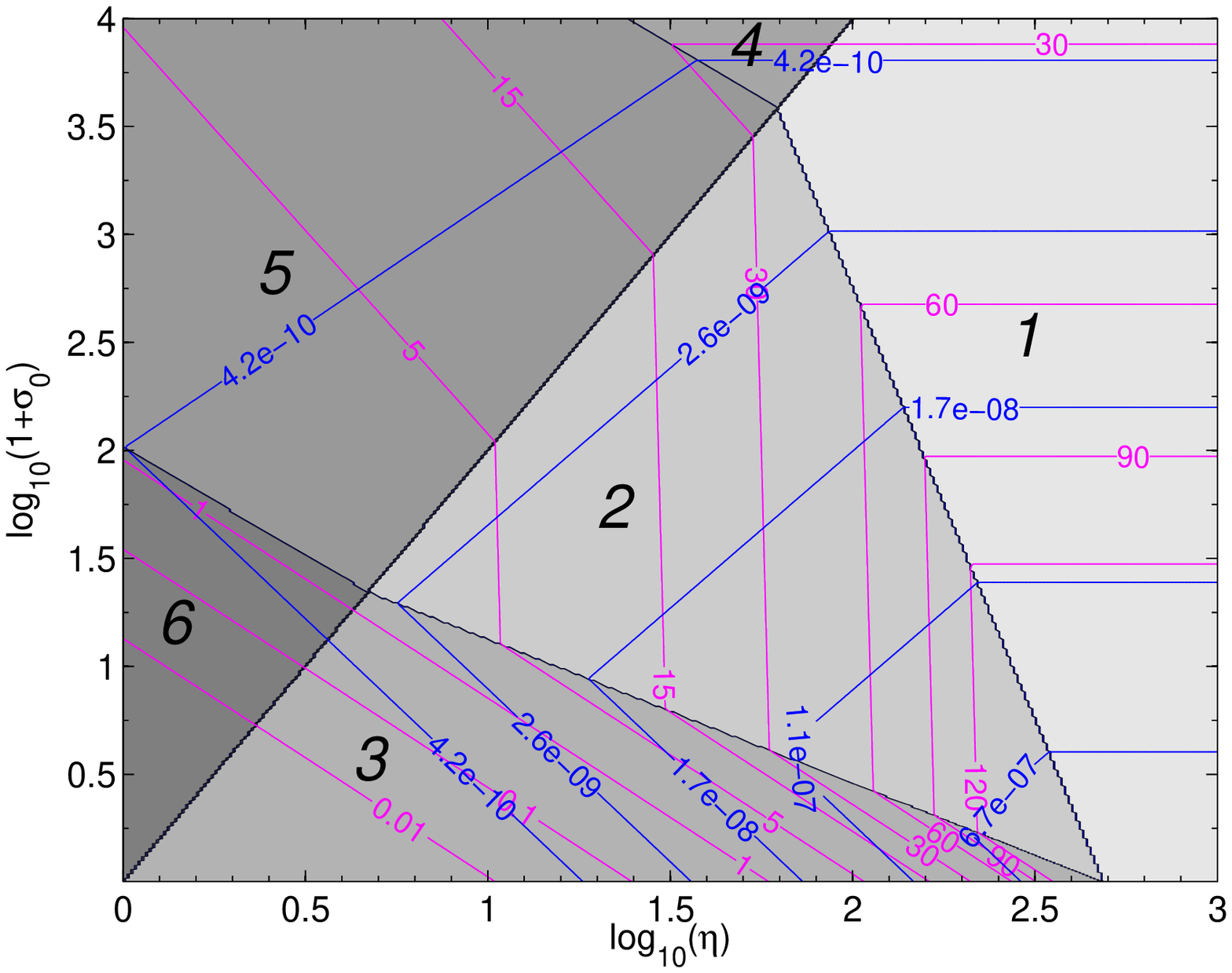}}
    \subfigure[]{
    \label{fig:subfig:c} %% label for first subfigure
    \includegraphics[width=3.0in]{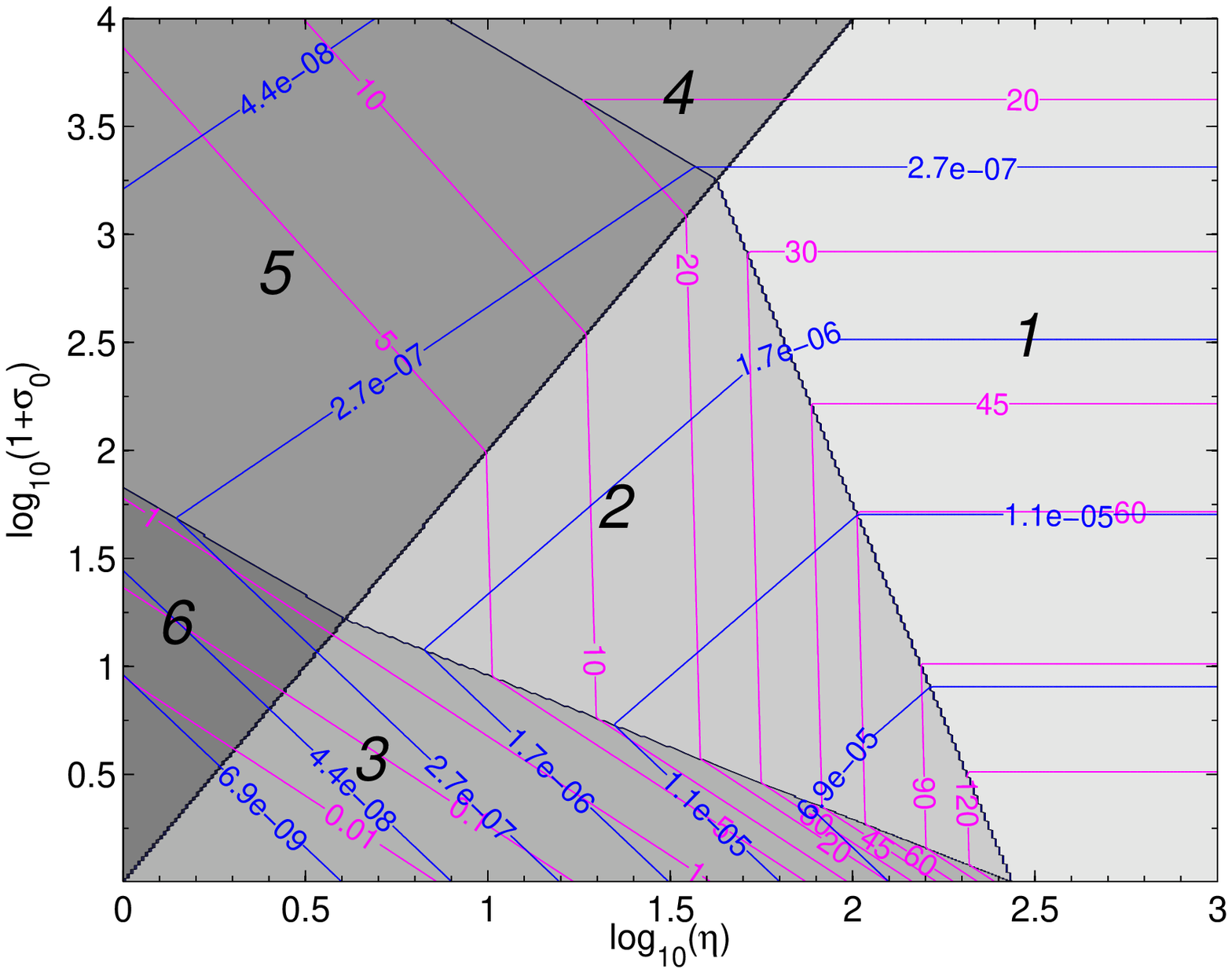}}
    \subfigure[]{
    \label{fig:subfig:d} %% label for first subfigure
    \includegraphics[width=3.0in]{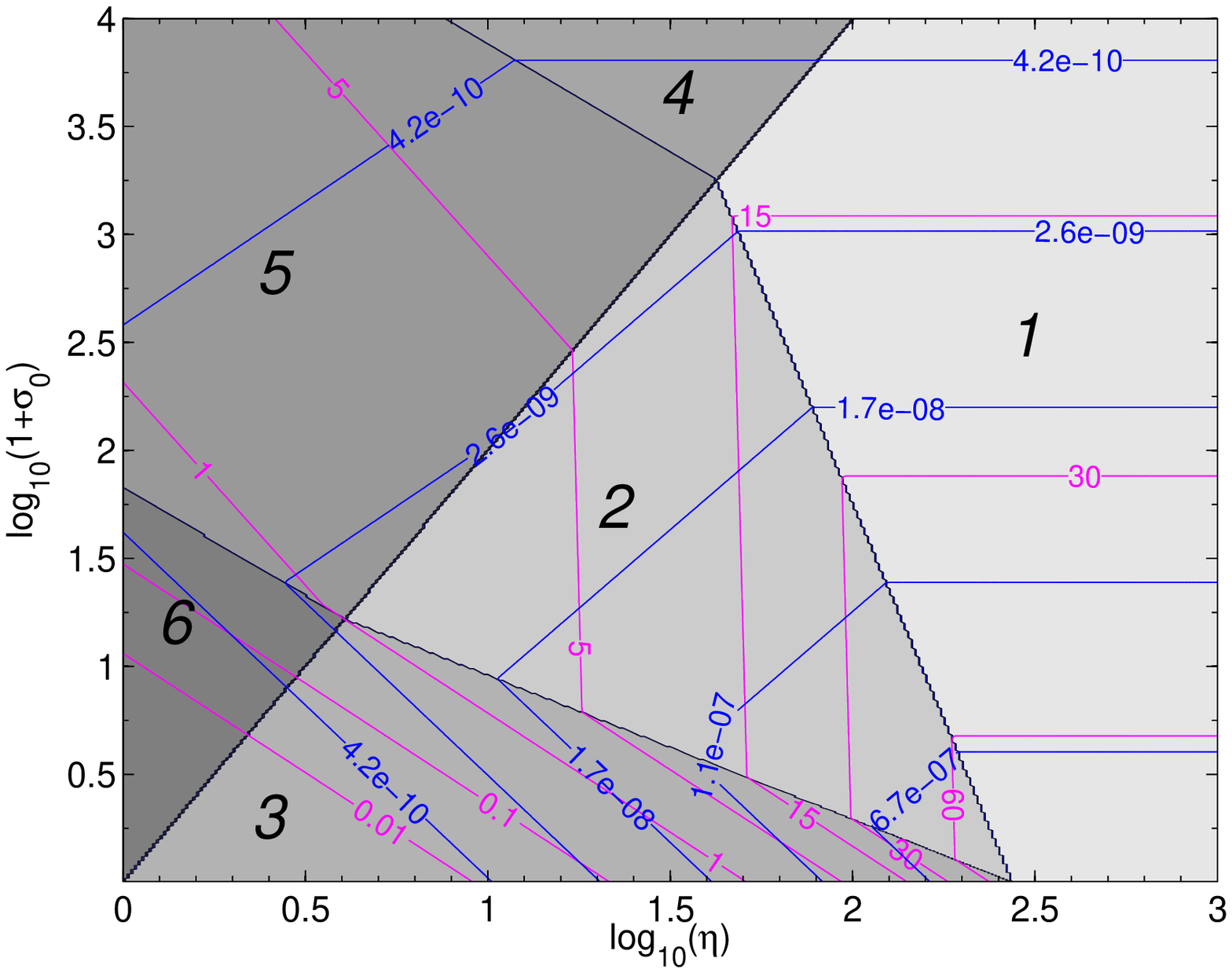}}
      \caption{Contour plots of $T_{\rm ob}$ and $F_{\rm BB}$ in the $(\eta, 1+\s)$ domain in the non-dissipative photosphere scenario. Pink lines are for $T_{\rm ob}$ in units of keV, and blue lines are for $F_{\rm BB}$ in units of ${\rm erg~cm^{-2}~ s^{-1}}$. The six regimes for the photosphere radius are shown in different grey blocks, with higher regime number ones marked with darker grey and the regime number marked in the block. For all the examples, $L_w=10^{52} {\rm erg~s^{-1}}$ is assumed. Top (bottom) panels are for $r_0=10^8 {\rm cm}$ ($r_0=10^{9} {\rm cm}$), respectively; and left (right) panels are for $z=0.1$ ($z=1$), respectively. }
           \label{fig:contour1}
            \end{figure}

It is useful to comment on the relationship between our work and several previous papers. \cite{pe'er07} were the first to propose to diagnose central engine parameters using the observed photosphere emission properties. Their method is within the standard fireball framework. Our general diagnostic method is consistent with \cite{pe'er07} when $\s_0 \ll 1$ is assumed. \cite{veres12} introduced the slow acceleration segment in a magnetically dominated flow to calculate the properties of the dissipative photosphere. Their photosphere radius is within the $\g \propto R^{1/3}$ regime (i.e. our regimes II and V). However, they did not introduce the $(1+\s_0)$ parameter to suppress the photosphere luminosity. This would over-estimate the photosphere brightness in more general cases. \cite{hascoet13} introduced a parameter to denote the fraction of thermal energy at the central engine, and calculated the brightness of the photosphere emission. However, they did not explicitly take into account the dynamical evolution of a high-$\sigma$ outflow. As shown in this paper, in order to reproduce the photosphere properties of the data (or upper limits), $\s_0$ is such that the coasting radius is usually above the IS radius. This suggests that ISs cannot operate efficiently, and significant jet energy dissipation has to proceed through magnetic dissipation processes such as ICMART. Based on the theoretical framework of \cite{hascoet13}, \cite{guiriec13} presented a method to infer $r_0$, $\eta$ and $r_{\rm ph}$ using observed quantities (their eqs.(3-5)). Their results agree with our regimes III and VI without magnetic dissipation (the coasting regime). They applied the method to the short GRB 120323A and obtained an anomalously low $\gamma$-ray efficiency. It is likely that the photosphere radius is in the slow magnetic acceleration regime, so that their diagnostic method should be replaced by ours in the regimes II and V. This would alleviate the low efficiency problem encountered by the burst. Finally, \cite{peng14} recently discovered a sub-dominant thermal component in the X-ray flares of some GRBs with a typical temperature of a few keV. According to Figures \ref{fig:contour1}, a low temperature, low flux thermal component typically requires a large $\s_0$. This is consistent with the theoretical argument that a magnetic mechanism has to play an important role to power an X-ray flare jet \citep{fan05}.

Finally, we'd like to point out several caveats of our approach. First, we have introduced a simple toy model to describe the dynamical evolution of a hybrid jet. We have assumed that the jet is accelerated intially thermally and later magnetically. Such a treatment was adopted by \cite{meszaros97}, and was proven to be valid for a specific hybrid MHD jet model \cite{vlahakis03}. We assume a linear acceleration below a ``rapid acceleration'' radius $r_{\rm ra}$ defined by the maximum of thermal coasting radius and the magneto-sonic point. We argue that this approximation is good for a hybrid jet. However, in reality the acceleration of a hybrid system in the rapid acceleration phase should be more complicated, requiring to solve a set of MHD equations with the contribution of a radiation force \cite[e.g.][]{russo13}. The solution may deviate from the simple linear acceleration assumption adopted here. If future detailed numerical simulations show deviation from this simple linear acceleration law below $r_{\rm ra}$, both of our bottom-up and top-down approaches should be modified accordingly. Another complication is the possible additional acceleration due to collimation of a stellar envelope \citep{tchekhovskoy09,tchekhovskoy10}. We argue that after the jet breaks out from the star, the confinement profile adopted by \cite{tchekhovskoy10}, which is valid during jet propagation inside the star \citep{zhangw03}, would be modified, so that the additional acceleration effect may not be significant. Our treatment neglected this effect. If it turns out that this effect is significant, then our treatment should be improved to include this correction effect.  In any case, since in most cases $r_{\rm ph}$ is above $r_{\rm ra}$, the approaches derived here should give correct results to order of magnitude. 

\begin{figure}
    \includegraphics[width=4.0in]{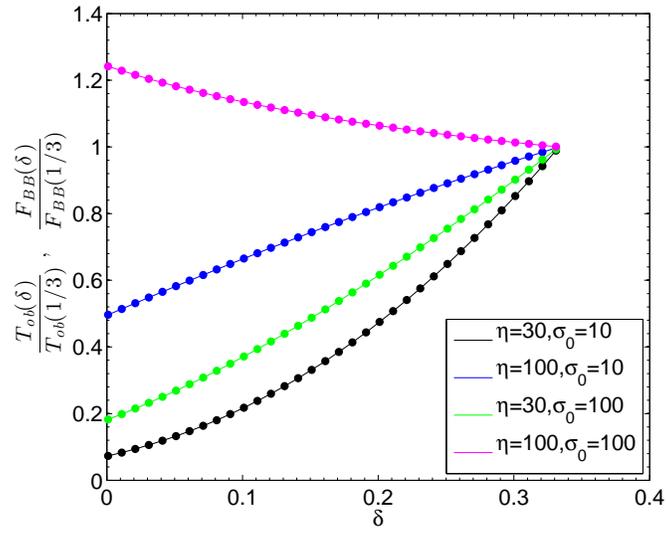} 
      \caption{Variation of $\frac{T_{\rm ob}(\delta)}{T_{\rm ob}(1/3)}$ (solid lines) and $\frac{F_{\rm BB}(\delta)}{F_{\rm BB}(1/3)}$ (dotted lines) as a function of $\delta$ in the non-dissipative photosphere models.}
           \label{fig:delta}
            \end{figure}

A final uncertainty of our approach is the acceleration index $\delta$ during the slow acceleration phase. For the bottom-up approach, we derived results for the arbitrary $\delta$ case (presented in the Appendix), but presented in the main text the $\delta = 1/3$ case. For the top-down approach, we only derived the formulae for the $\delta=1/3$ case. For the non-dissipative photosphere model we focus in the paper, such an acceleration law may be achieved for an impulsive jet, as is the case for GRBs \citep{granot11}. Nonetheless, a shallower acceleration law with $\delta < 1/3$ may be possible, which would lead to corrections to the results from the $\delta=1/3$ model. In Figure \ref{fig:delta}, we show how the photosphere properties vary with $\delta$. It turns out that the ratio $T_{\rm ob}(\delta) / T_{\rm ob}(1/3)$ is the same as the ratio $F_{\rm BB}(\delta) / F_{\rm BB}(1/3)$, which varies by a factor of $50\%$ as long as $\delta$ is not too small (say, above 0.2). In most cases that are relevant to GRBs (e.g. $100< \eta (1+\sigma_0) < 10000$), the temperature and flux derived with $\delta=1/3$ set an upper limit for the more general $\delta$ models. In any case, with $\delta < 1/3$, magnetic acceleration is less efficient during the slow acceleration regime, so that it is even more difficult to reach the coasting phase before deceleration. The parameter space for internal shocks to operate is further reduced, and it is more likely that ICMART is the main mechanism to power bright non-thermal emission from GRBs.

\acknowledgements 
We thank Peter M\'esz\'aros, Asaf Pe'er, Robert Mochkovitch, and Z. Lucas Uhm for helpful comments, and an anonymous referee for a constructive report. This work is partially supported by NASA through grant NNX14AF85G.

\clearpage

\appendix{Appendix}

In this Appendix, we present the results for an arbitrary $\delta$ value. The corresponding expressions in regimes II and V in Section \ref{sec:no-mag-bottom} read the following:

Regime II:

\begin{eqnarray}
&&r_{\rm ra}=1.0\times 10^{11}~{\rm cm}~r_{0,9}\eta_2 ,\nonumber\\
&&r_{\rm c}=1.0\times 10^{\frac{11\delta+2}{\delta}}~{\rm cm}~r_{0,9}\eta_2(1+\s_0)_2^{\frac{1}{\delta}},\nonumber\\
&&r_{\rm ph}=5.8\times 10^{\frac{22\delta+10}{2\delta+1}}~{\rm cm}~L_{w,52}^{\frac{1}{2\delta+1}}r_{0,9}^{\frac{2\delta}{2\delta+1}}\eta_2^{\frac{2\delta-3}{2\delta+1}}(1+\s_0)_2^{-\frac{1}{2\delta+1}},\nonumber\\
&&\g_{\rm ph}=5.8\times 10^{\frac{3\delta+2}{2\delta+1}}~L_{w,52}^{\frac{\delta}{2\delta+1}}r_{0,9}^{-\frac{\delta}{2\delta+1}}\eta_2^{\frac{1-2\delta}{2\delta+1}}(1+\s_0)_2^{-\frac{\delta}{2\delta+1}},\nonumber\\
&&1+\s_{\rm ph}=10^{\frac{19\delta+2}{2\delta+1}}e^{-\frac{43.2\delta}{2\delta+1}}L_{w,52}^{-\frac{\delta}{2\delta+1}}r_{0,9}^{\frac{\delta}{2\delta+1}}\eta_2^{\frac{4\delta}{2\delta+1}}(1+\s_0)_2^{\frac{3\delta+1}{2\delta+1}},\nonumber\\
&&k T_{\rm ob}=1.2\times 10^{\frac{43-28\delta}{6\delta+3}}e^{\frac{86.4\delta-86.4}{6\delta+3}}~{\rm keV}~(1+z)^{-1}L_{w,52}^{\frac{14\delta-5}{24\delta+12}}r_{0,9}^{\frac{1-10\delta}{12\delta+6}}\eta_2^{\frac{8-8\delta}{6\delta+3}}(1+\s_0)_2^{\frac{5-14\delta}{24\delta+12}},\nonumber\\
&&F_{\rm BB}=2.3\times 10^{\frac{16-82\delta}{6\delta+3}}e^{\frac{86.4\delta-86.4}{6\delta+3}}~{\rm erg~s^{-1}cm^{-2}}~L_{w,52}^{\frac{8\delta+1}{6\delta+3}}r_{0,9}^{\frac{2-2\delta}{6\delta+3}}\eta_2^{\frac{8-8\delta}{6\delta+3}}(1+\s_0)_2^{-\frac{8\delta+1}{6\delta+3}}d_{L,28}^{-2}.\nonumber\\
\end{eqnarray}

Regime V:

\begin{eqnarray}
&&r_{\rm ra}=2.15\times 10^{10}~{\rm cm}~r_{0,9}\eta_2^{1/3}(1+\s_0)_2^{1/3},\nonumber\\
&&r_{\rm c}=2.15\times 10^{\frac{31\delta+8}{3\delta}}~{\rm cm}~r_{0,9}\eta_2^{\frac{\delta+2}{3\delta}}(1+\s_0)_2^{\frac{\delta+2}{3\delta}},\nonumber\\
&&r_{\rm ph}=5.8\times 10^{\frac{62\delta+34}{6\delta+3}}~{\rm cm}~L_{w,52}^{\frac{1}{2\delta+1}}r_{0,9}^{\frac{2\delta}{2\delta+1}}\eta_2^{\frac{2\delta-5}{6\delta+3}}(1+\s_0)_2^{\frac{2\delta-5}{6\delta+3}},\nonumber\\
&&\g_{\rm ph}=5.8\times 10^{\frac{11\delta+4}{6\delta+3}}~L_{w,52}^{\frac{\delta}{2\delta+1}}r_{0,9}^{-\frac{\delta}{2\delta+1}}\eta_2^{\frac{1-4\delta}{6\delta+3}}(1+\s_0)_2^{\frac{1-4\delta}{6\delta+3}},\nonumber\\
&&1+\s_{\rm ph}=10^{\frac{19\delta+2}{2\delta+1}}e^{-\frac{43.2\delta}{2\delta+1}}L_{w,52}^{-\frac{\delta}{2\delta+1}}r_{0,9}^{\frac{\delta}{2\delta+1}}\eta_2^{\frac{10\delta+2}{6\delta+3}}(1+\s_0)_2^{\frac{10\delta+2}{6\delta+3}},\nonumber\\
&&k T_{\rm ob}=4.3\times 10^{\frac{112-82\delta}{18\delta+9}}e^{\frac{86.4\delta-86.4}{6\delta+3}}~{\rm keV}~(1+z)^{-1}L_{w,52}^{\frac{14\delta-5}{24\delta+12}}r_{0,9}^{\frac{1-10\delta}{12\delta+6}}\eta_2^{\frac{4-4\delta}{6\delta+3}}(1+\s_0)_2^{\frac{13-22\delta}{24\delta+12}},\nonumber\\
&&F_{\rm BB}=0.83\times 10^{\frac{40-226\delta}{18\delta+9}}e^{\frac{86.4\delta-86.4}{6\delta+3}}~{\rm erg~s^{-1}cm^{-2}}~L_{w,52}^{\frac{8\delta+1}{6\delta+3}}r_{0,9}^{\frac{2-2\delta}{6\delta+3}}\eta_2^{\frac{4-4\delta}{6\delta+3}}(1+\s_0)_2^{\frac{1-10\delta}{6\delta+3}}d_{L,28}^{-2}.\nonumber\\
\end{eqnarray}

Similarly, the corresponding expressions for Regimes II, III, V, and VI in Section \ref{sec:mag-bottom} read the following:

Regime II
\begin{eqnarray}
&&r_{\rm ra}=1.0\times 10^{11}~{\rm cm}~r_{0,9}\eta_2 ,\nonumber\\
&&r_{\rm c}=1.0\times 10^{\frac{11\delta+2}{\delta}}~{\rm cm}~r_{0,9}\eta_2(1+\s_0)_2^{\frac{1}{\delta}},\nonumber\\
&&r_{\rm ph}=5.8\times 10^{\frac{22\delta+10}{2\delta+1}}~{\rm cm}~L_{w,52}^{\frac{1}{2\delta+1}}r_{0,9}^{\frac{2\delta}{2\delta+1}}\eta_2^{\frac{2\delta-3}{2\delta+1}}(1+\s_0)_2^{-\frac{1}{2\delta+1}},\nonumber\\
&&\g_{\rm ph}=5.8\times 10^{\frac{3\delta+2}{2\delta+1}}~L_{w,52}^{\frac{\delta}{2\delta+1}}r_{0,9}^{-\frac{\delta}{2\delta+1}}\eta_2^{\frac{1-2\delta}{2\delta+1}}(1+\s_0)_2^{-\frac{\delta}{2\delta+1}},\nonumber\\
&&1+\s_{\rm ph}=10^{\frac{19\delta+2}{2\delta+1}}e^{-\frac{43.2\delta}{2\delta+1}}L_{w,52}^{-\frac{\delta}{2\delta+1}}r_{0,9}^{\frac{\delta}{2\delta+1}}\eta_2^{\frac{4\delta}{2\delta+1}}(1+\s_0)_2^{\frac{3\delta+1}{2\delta+1}},\nonumber\\
&&k T_{\rm BB}^{\rm eff}=1.8\times 10^{\frac{44-45\delta}{8\delta+4}}e^{\frac{129.6\delta-86.4}{8\delta+4}}~{\rm keV}~(1+z)^{-1}L_{w,52}^{\frac{5\delta-1}{8\delta+4}}r_{0,9}^{-\frac{7\delta}{8\delta+4}}\eta_2^{\frac{2-3\delta}{2\delta+1}}(1+\s_0)_2^{\frac{1-5\delta}{8\delta+4}},\nonumber\\
&&F_{\rm ph}=5.8\times 10^{-\frac{15\delta+7}{2\delta+1}}~{\rm erg~s^{-1}cm^{-2}}~L_{w,52}^{\frac{3\delta+1}{2\delta+1}}r_{0,9}^{-\frac{\delta}{2\delta+1}}\eta_2^{-\frac{4\delta}{2\delta+1}}(1+\s_0)_2^{-\frac{3\delta+1}{2\delta+1}}d_{L,28}^{-2}.\nonumber\\
\end{eqnarray}

Regime III:

\begin{eqnarray}
&&r_{\rm ra}=1.0\times 10^{11}~{\rm cm}~r_{0,9}\eta_2 ,\nonumber\\
&&r_{\rm c}=1.0\times 10^{\frac{11\delta+2}{\delta}}~{\rm cm}~r_{0,9}\eta_2(1+\s_0)_2^{\frac{1}{\delta}},\nonumber\\
&&r_{\rm ph}=5.81\times 10^{12}~{\rm cm}~ L_{w,52}\eta_1^{-3}(1+\s_0)_1^{-3},\nonumber\\
&&\g_{\rm ph}=100 \eta_1(1+\s_0)_1,\nonumber\\
&&1+\s_{\rm ph} \simeq 1,\nonumber\\
&&k T_{\rm BB}^{\rm eff}=8.1\times10^{\frac{1}{6\delta}}~{\rm keV}~(1+z)^{-1}L_{w,52}^{-5/12}r_{0,9}^{1/6}\eta_1^{8/3}(1+\s_0)_1^{\frac{15\delta+1}{6\delta}},\nonumber\\
&&F_{\rm ph}=1.54\times 10^{\frac{2-21\delta}{3\delta}}~{\rm erg~s^{-1}cm^{-2}}~ L_{w,52}^{1/3}r_{0,9}^{2/3}\eta_1^{8/3}(1+\s_0)_1^{\frac{6\delta+2}{3\delta}}d_{L,28}^{-2}.\nonumber\\
\end{eqnarray}

Regime V:

\begin{eqnarray}
&&r_{\rm ra}=2.15\times 10^{10}~{\rm cm}~r_{0,9}\eta_2^{1/3}(1+\s_0)_2^{1/3},\nonumber\\
&&r_{\rm c}=1.0\times 10^{\frac{31\delta+8}{3\delta}}~{\rm cm}~r_{0,9}\eta_2^{\frac{\delta+2}{3\delta}}(1+\s_0)_2^{\frac{\delta+2}{3\delta}},\nonumber\\
&&r_{\rm ph}=5.8\times 10^{\frac{62\delta+34}{6\delta+3}}~{\rm cm}~L_{w,52}^{\frac{1}{2\delta+1}}r_{0,9}^{\frac{2\delta}{2\delta+1}}\eta_2^{\frac{2\delta-5}{6\delta+3}}(1+\s_0)_2^{\frac{2\delta-5}{6\delta+3}},\nonumber\\
&&\g_{\rm ph}=5.8\times 10^{\frac{11\delta+4}{6\delta+3}}~L_{w,52}^{\frac{\delta}{2\delta+1}}r_{0,9}^{-\frac{\delta}{2\delta+1}}\eta_2^{\frac{1-4\delta}{6\delta+3}}(1+\s_0)_2^{\frac{1-4\delta}{6\delta+3}},\nonumber\\
&&1+\s_{\rm ph}=10^{\frac{19\delta+2}{2\delta+1}}e^{-\frac{43.2\delta}{2\delta+1}}L_{w,52}^{-\frac{\delta}{2\delta+1}}r_{0,9}^{\frac{\delta}{2\delta+1}}\eta_2^{\frac{10\delta+2}{6\delta+3}}(1+\s_0)_2^{\frac{10\delta+2}{6\delta+3}},\nonumber\\
&&k T_{\rm BB}^{\rm eff}=8.2\times 10^{\frac{112-133\delta}{24\delta+12}}e^{\frac{129.6\delta-86.4}{8\delta+4}}~{\rm keV}~(1+z)^{-1}L_{w,52}^{\frac{5\delta-1}{8\delta+4}}r_{0,9}^{-\frac{7\delta}{8\delta+4}}\eta_2^{\frac{2-3\delta}{4\delta+2}}(1+\s_0)_2^{\frac{3-8\delta}{8\delta+4}},\nonumber\\
&&F_{\rm ph}=5.8\times 10^{-\frac{13\delta+7}{2\delta+1}}~{\rm erg~s^{-1}cm^{-2}}~L_{w,52}^{\frac{3\delta+1}{2\delta+1}}r_{0,9}^{-\frac{\delta}{2\delta+1}}\eta_2^{-\frac{2\delta}{2\delta+1}}(1+\s_0)_2^{-\frac{4\delta+1}{2\delta+1}}d_{L,28}^{-2}.\nonumber\\
\end{eqnarray}

Regime VI:

\begin{eqnarray}
&&r_{\rm ra}=2.15\times 10^{10}~{\rm cm}~r_{0,9}\eta_2^{1/3}(1+\s_0)_2^{1/3},\nonumber\\
&&r_{\rm c}=1.0\times 10^{\frac{31\delta+8}{3\delta}}~{\rm cm}~r_{0,9}\eta_2^{\frac{\delta+2}{3\delta}}(1+\s_0)_2^{\frac{\delta+2}{3\delta}},\nonumber\\
&&r_{\rm ph}=5.81\times 10^{12}~{\rm cm}~ L_{w,52}\eta_1^{-3}(1+\s_0)_1^{-3},\nonumber\\
&&1+\g_{\rm ph}=100 \eta_1(1+\s_0)_1,\nonumber\\
&&1+\s_{\rm ph} \simeq 1,\nonumber\\
&&k T_{\rm BB}^{\rm eff}=8.6\times 10^{\frac{2}{9\delta}}~{\rm keV}~(1+z)^{-1}L_{w,52}^{-5/12}r_{0,9}^{1/6}\eta_1^{\frac{49\delta+2}{18\delta}}(1+\s_0)_1^{\frac{89\delta+4}{36\delta}},\nonumber\\
&&F_{\rm ph}=2.0\times 10^{\frac{8-63\delta}{9\delta}}~{\rm erg~s^{-1}cm^{-2}}~L_{w,52}^{1/3}r_{0,9}^{2/3}\eta_1^{\frac{26\delta+4}{9\delta}}(1+\s_0)_1^{\frac{17\delta+4}{9\delta}}d_{L,28}^{-2}.\nonumber\\
\end{eqnarray}

For the general $\delta$ models, the criteria of all the twelve regimes based on the central engine properties are collected in Table 3.

\clearpage

\begin{sidewaystable}
\label{table:regime1}
\caption{Definition and theoretical criteria of $r_{\rm ph}$ regimes for different models, with $\delta=1/3$.}
\begin{tabular}{llllllll}
\hline\hline

           &~~~~~~~~~~~~~~~~~~~$r_{\rm ph}<r_{\rm ra}$                                                         &~~~~~~~~~~~~~~~~~~~~$r_{\rm ra}<r_{\rm ph}<r_{\rm c}$                                                                  &~~~~~~~~~~~~~~~~~~~~$r_{\rm ph}>r_{\rm c}$                                                                   \\

\hline
 Non-dissipation           &~~~~~~~~~~~~~~~~~~~  Regime I                                       &   ~~~~~~~~~~~~~~~~~~~~~~Regime II &~~~~~~~~~~~~~~~~~~~  Regime III\\
 \hline
   &  $\eta^{12/5}(1+\s_0)^{3/5}>7.22\times 10^{5}L_{w,52}^{3/5}r_{0,9}^{-3/5}$                                       &  $\eta^{12/5}(1+\s_0)^{3/5}<7.22\times 10^{5}L_{w,52}^{3/5}r_{0,9}^{-3/5}$&  $\eta^{12/5}(1+\s_0)^{18/5}<7.22\times 10^{5}L_{w,52}^{3/5}r_{0,9}^{-3/5}$
                    \\
 $\eta>(1+\s_0)^{1/2}$       &  $ ~$                                        &  $\eta^{12/5}(1+\s_0)^{18/5}>7.22\times 10^{5}L_{w,52}^{3/5}r_{0,9}^{-3/5}$        &  $~$                 \\
\hline
Non-dissipation     &~~~~~~~~~~~~~~~~~~~  Regime IV                                       &   ~~~~~~~~~~~~~~~~~~~~~~Regime V &~~~~~~~~~~~~~~~~~~~  Regime VI\\
 \hline     & $\eta^{6/5}(1+\s_0)^{6/5}>7.22\times 10^{5}L_{w,52}^{3/5}r_{0,9}^{-3/5}
$  &  $\eta^{6/5}(1+\s_0)^{6/5}<7.22\times 10^{5}L_{w,52}^{3/5}r_{0,9}^{-3/5}
$                  &  $\eta^{16/5}(1+\s_0)^{16/5}<7.22\times 10^{5}L_{w,52}^{3/5}r_{0,9}^{-3/5}$    \\
  $\eta<(1+\s_0)^{1/2}$        &  $~$                                        &  $\eta^{16/5}(1+\s_0)^{16/5}>7.22\times 10^{5}L_{w,52}^{3/5}r_{0,9}^{-3/5}$                        &  $~$ \\
\hline
 Dissipation   &~~~~~~~~~~~~~~~~~~~  Regime I                                       &   ~~~~~~~~~~~~~~~~~~~~~~Regime II &~~~~~~~~~~~~~~~~~~~  Regime III\\
 \hline       &  $\eta^{12/5}(1+\s_0)^{3/5}>7.22\times 10^{5}L_{w,52}^{3/5}r_{0,9}^{-3/5}$                                             &  $\eta^{12/5}(1+\s_0)^{3/5}<7.22\times 10^{5}L_{w,52}^{3/5}r_{0,9}^{-3/5}$         &  $\eta^{12/5}(1+\s_0)^{18/5}<7.22\times 10^{5} L_{w,52}^{3/5}r_{0,9}^{-3/5}$
          \\
  $\eta>(1+\s_0)^{1/2}$        &  $~$                                        &  $\eta^{12/5}(1+\s_0)^{18/5}>7.22\times 10^{5} L_{w,52}^{3/5}r_{0,9}^{-3/5}$        &  $~$                 \\
\hline
Dissipation    &~~~~~~~~~~~~~~~~~~~  Regime IV                                       &   ~~~~~~~~~~~~~~~~~~~~~~Regime V &~~~~~~~~~~~~~~~~~~~  Regime VI\\
 \hline      &  $\eta^{6/5}(1+\s_0)^{6/5}>7.22\times 10^{5}L_{w,52}^{3/5}r_{0,9}^{-3/5}$                                            &  $\eta^{6/5}(1+\s_0)^{6/5}<7.22\times 10^{5}L_{w,52}^{3/5}r_{0,9}^{-3/5}$      &  $\eta^{16/5}(1+\s_0)^{16/5}<7.22\times 10^{5}L_{w,52}^{3/5}r_{0,9}^{-3/5}$              \\
  $\eta<(1+\s_0)^{1/2}$        &  $~$                                        &  $\eta^{16/5}(1+\s_0)^{16/5}>7.22\times 10^{5} L_{w,52}^{3/5}r_{0,9}^{-3/5}$                        &  $~$ \\
  \hline
\end{tabular}
\end{sidewaystable}

\begin{table}
\label{table:regime2}
\caption{Observational criteria of $r_{\rm ph}$ regimes for different models.}
\begin{tabular}{llllllll}
\hline\hline
             No dissipation Regime II                                                                         \\
\hline
 $14.8(1+z)^{1/4}(\frac{k T_{\rm ob}}{50~{\rm keV}})^{1/4}(\frac{F_{\rm BB}}{10^{-8}~{\rm erg~s^{-1}cm^{-2}}})^{3/16}r_{0,9}^{-1/8}f_{\rm th,-1}^{1/2}f_{\gamma}^{1/2}d_{L,28}^{3/8}>1$  \\
$0.24(1+z)^{-3}(\frac{k T_{\rm ob}}{50~{\rm keV}})^{-3}(\frac{F_{\rm BB}}{10^{-8}~{\rm erg~s^{-1}cm^{-2}}})^{3/4}r_{0,9}^{-3/2}d_{L,28}^{3/2}>1$ \\
$1.43\times10^{-5}(1+z)^{-7}(\frac{k T_{\rm ob}}{50~{\rm keV}})^{-7}(\frac{F_{\rm BB}}{10^{-8}~{\rm erg~s^{-1}cm^{-2}}})^{7/4}r_{0,9}^{-7/2}f_{\rm th,-1}^{3}f_{\gamma}^{3}d_{L,28}^{7/2}<1$\\
\hline
           No dissipation Regime III \\
\hline
$8.28(1+z)^{-3/2}(\frac{k T_{\rm ob}}{30~{\rm keV}})^{-3/2}(\frac{F_{\rm BB}}{10^{-7}~{\rm erg~s^{-1}cm^{-2}}})^{5/8}r_{0,9}^{-1}f_{\rm th,-1}^{5/4}f_{\gamma}^{5/4}d_{L,28}^{5/4}>1$                  \\
$9.42\times 10^{-2}(1+z)^{-14/3}(\frac{k T_{\rm ob}}{30~{\rm keV}})^{-14/3}(\frac{F_{\rm BB}}{10^{-7}~{\rm erg~s^{-1}cm^{-2}}})^{7/6}r_{0,9}^{-7/3}f_{\rm th,-1}^{2}f_{\gamma}^{2}d_{L,28}^{7/3}>1$   \\ 
\hline
Non-dissipation       Regime V  \\
\hline
$41.4(1+z)^{1/2}(\frac{k T_{\rm ob}}{10~{\rm keV}})^{1/2}(\frac{F_{\rm BB}}{10^{-9}~{\rm erg~s^{-1}cm^{-2}}})^{3/8}r_{0,9}^{-1/4}f_{\rm th,-1}f_{\gamma}d_{L,28}^{3/4}<1$\\
$5.28(1+z)^{-3}(\frac{k T_{\rm ob}}{10~{\rm keV}})^{-3}(\frac{F_{\rm BB}}{10^{-9}~{\rm erg~s^{-1}cm^{-2}}})^{3/4}r_{0,9}^{-3/2}d_{L,28}^{3/2}>1$\\
$1.16\times 10^{-5}(1+z)^{-8}(\frac{k T_{\rm ob}}{10~{\rm keV}})^{-8}(\frac{F_{\rm BB}}{10^{-9}~{\rm erg~s^{-1}cm^{-2}}})r_{0,9}^{-3}f_{\rm th,-1}f_{\gamma}d_{L,28}^{2}<1$\\
\hline
Non-dissipation       Regime VI \\
\hline
$8.28(1+z)^{-3/2}(\frac{k T_{\rm ob}}{30~{\rm keV}})^{-3/2}(\frac{F_{\rm BB}}{10^{-7}~{\rm erg~s^{-1}cm^{-2}}})^{5/8}r_{0,9}^{-1}f_{\rm th,-1}^{5/4}f_{\gamma}^{5/4}d_{L,28}^{5/4}<1$                  \\
 $5.63\times 10^{-3}(1+z)^{-8/3}(\frac{k T_{\rm ob}}{30~{\rm keV}})^{-8/3}(\frac{F_{\rm BB}}{10^{-7}~{\rm erg~s^{-1}cm^{-2}}})^{1/3}r_{0,9}^{-1}f_{\rm th,-1}^{1/3}f_{\gamma}^{1/3}d_{L,28}^{2/3}>1$              \\ 
\hline
\end{tabular}
\end{table}

\begin{sidewaystable}
\label{table:regime1}
\caption{Definition and theoretical criteria of $r_{\rm ph}$ regimes for different models, with general $\delta$ value.}
\begin{tabular}{llllllll}
\hline\hline

           &~~~~~~~~~~~~~~~~~~~$r_{\rm ph}<r_{\rm ra}$                                                         &~~~~~~~~~~~~~~~~~~~~$r_{\rm ra}<r_{\rm ph}<r_{\rm c}$                                                                  &~~~~~~~~~~~~~~~~~~~~$r_{\rm ph}>r_{\rm c}$                                                                   \\

\hline
 Non-dissipation           &~~~~~~~~~~~~~~~~~~~  Regime I                                       &   ~~~~~~~~~~~~~~~~~~~~~~Regime II &~~~~~~~~~~~~~~~~~~~  Regime III\\
 \hline
   &  $\eta^{12/5}(1+\s_0)^{3/5}>7.22\times 10^{5}L_{w,52}^{3/5}r_{0,9}^{-3/5}$                                       &  $\eta^{\frac{4}{2\delta+1}}(1+\s_0)^{\frac{1}{2\delta+1}}<5.81\times 10^{\frac{9}{2\delta+1}}L_{w,52}^{\frac{1}{2\delta+1}}r_{0,9}^{-\frac{1}{2\delta+1}}$&  $\eta^{12/5}(1+\s_0)^{\frac{9\delta+3}{5\delta}}<7.22\times 10^{5}L_{w,52}^{3/5}r_{0,9}^{-3/5}$
                    \\
 $\eta>(1+\s_0)^{1/2}$       &  $ ~$                                        &  $\eta^{\frac{4}{2\delta+1}}(1+\s_0)^{\frac{3\delta+1}{\delta(2\delta+1)}}>5.81\times 10^{\frac{9}{2\delta+1}}L_{w,52}^{\frac{1}{2\delta+1}}r_{0,9}^{-\frac{1}{2\delta+1}}$       &  $~$                 \\
\hline
Non-dissipation     &~~~~~~~~~~~~~~~~~~~  Regime IV                                       &   ~~~~~~~~~~~~~~~~~~~~~~Regime V &~~~~~~~~~~~~~~~~~~~  Regime VI\\
 \hline     & $\eta^{6/5}(1+\s_0)^{6/5}>7.22\times 10^{5}L_{w,52}^{3/5}r_{0,9}^{-3/5}
$  &  $\eta^{\frac{2}{2\delta+1}}(1+\s_0)^{\frac{2}{2\delta+1}}<5.81\times 10^{\frac{9}{2\delta+1}}L_{w,52}^{\frac{1}{2\delta+1}}r_{0,9}^{-\frac{1}{2\delta+1}}$                  &  $\eta^{\frac{10\delta+2}{5\delta}}(1+\s_0)^{\frac{10\delta+2}{5\delta}}<7.22\times 10^{5}L_{w,52}^{3/5}r_{0,9}^{-3/5}$    \\
  $\eta<(1+\s_0)^{1/2}$        &  $~$                                        &  $\eta^{\frac{10\delta+2}{3\delta(2\delta+1)}}(1+\s_0)^{\frac{10\delta+2}{3\delta(2\delta+1)}}>5.81\times 10^{\frac{9}{2\delta+1}}L_{w,52}^{\frac{1}{2\delta+1}}r_{0,9}^{-\frac{1}{2\delta+1}}$                       &  $~$ \\
\hline
 Dissipation   &~~~~~~~~~~~~~~~~~~~  Regime I                                       &   ~~~~~~~~~~~~~~~~~~~~~~Regime II &~~~~~~~~~~~~~~~~~~~  Regime III\\
 \hline       &  $\eta^{12/5}(1+\s_0)^{3/5}>7.22\times 10^{5}L_{w,52}^{3/5}r_{0,9}^{-3/5}$                                             &  $\eta^{\frac{4}{2\delta+1}}(1+\s_0)^{\frac{1}{2\delta+1}}<5.81\times 10^{\frac{9}{2\delta+1}}L_{w,52}^{\frac{1}{2\delta+1}}r_{0,9}^{-\frac{1}{2\delta+1}}$         &  $\eta^{12/5}(1+\s_0)^{\frac{9\delta+3}{5\delta}}<7.22\times 10^{5}L_{w,52}^{3/5}r_{0,9}^{-3/5}$
          \\
  $\eta>(1+\s_0)^{1/2}$        &  $~$                                        &  $\eta^{\frac{4}{2\delta+1}}(1+\s_0)^{\frac{3\delta+1}{\delta(2\delta+1)}}>5.81\times 10^{\frac{9}{2\delta+1}}L_{w,52}^{\frac{1}{2\delta+1}}r_{0,9}^{-\frac{1}{2\delta+1}}$        &  $~$                 \\
\hline
Dissipation    &~~~~~~~~~~~~~~~~~~~  Regime IV                                       &   ~~~~~~~~~~~~~~~~~~~~~~Regime V &~~~~~~~~~~~~~~~~~~~  Regime VI\\
 \hline      &  $\eta^{6/5}(1+\s_0)^{6/5}>7.22\times 10^{5}L_{w,52}^{3/5}r_{0,9}^{-3/5}$                                            &  $\eta^{\frac{2}{2\delta+1}}(1+\s_0)^{\frac{2}{2\delta+1}}<5.81\times 10^{\frac{9}{2\delta+1}}L_{w,52}^{\frac{1}{2\delta+1}}r_{0,9}^{-\frac{1}{2\delta+1}}$      &  $\eta^{\frac{10\delta+2}{5\delta}}(1+\s_0)^{\frac{10\delta+2}{5\delta}}<7.22\times 10^{5}L_{w,52}^{3/5}r_{0,9}^{-3/5}$              \\
  $\eta<(1+\s_0)^{1/2}$        &  $~$                                        &  $\eta^{\frac{10\delta+2}{3\delta(2\delta+1)}}(1+\s_0)^{\frac{10\delta+2}{3\delta(2\delta+1)}}>5.81\times 10^{\frac{9}{2\delta+1}}L_{w,52}^{\frac{1}{2\delta+1}}r_{0,9}^{-\frac{1}{2\delta+1}}$                       &  $~$ \\
  \hline
\end{tabular}
\end{sidewaystable}

\clearpage

\clearpage

     \clearpage
            
\clearpage

\end{document}